%% file: OrionBar_pyPAHdb_R1.tex
\documentclass[longauth]{aa}  

\makeatletter
\renewcommand*\aa@pageof{, page \thepage{} of \pageref*{LastPage}}
\makeatother
\usepackage{graphicx}
\graphicspath{{figures/}}
\usepackage{soul}
\usepackage{xcolor}
\usepackage{txfonts}
\usepackage[breaklinks=true, hidelinks,colorlinks=true,linkcolor=blue,citecolor=blue]{hyperref}
\usepackage{orcidlink}

\newcommand{\HII}[1]{H$\,\textsc{ii}$}
\newcommand{\Nc}[1]{N$_{\textrm{C}}$}
\newcommand{\ANc}[1]{$\overline{N_C}$}
\newcommand{\micron}[1]{$\mu$m}

\input{affiliations}

\begin{document}

   \title{PDRs4All\\ XVIII. The evolution of the PAH ionisation and PAH size distribution across the Orion Bar}

   \author{Alexandros Maragkoudakis \inst{\ames} \orcidlink{0000-0003-2552-3871} \and
            Christiaan Boersma \inst{\ames} \orcidlink{0000-0002-4836-217X} \and
            Els Peeters \inst{\uwo, \wspace, \seti} \orcidlink{0000-0002-2541-1602} \and
            Louis J. Allamandola \inst{\ames} \orcidlink{0000-0002-6049-4079} \and
            Pasquale Temi \inst{\ames} \orcidlink{0000-0002-8341-342X} \and
            Vincent J. Esposito \inst{\ames} \orcidlink{0000-0001-6035-3869} \and
            Jesse D. Bregman \inst{\ames} \orcidlink{0000-0002-1440-5362} \and
            Alessandra Ricca \inst{\ames, \seti} \orcidlink{0000-0002-3141-0630} \and
            Felipe Alarc\'on \inst{\umich} \orcidlink{0000-0002-2692-7862} \and
            Olivier Bern\'{e} \inst{\toulouse} \orcidlink{0000-0002-1686-8395} \and
            Mridusmita Buragohain \inst{\uoh} \orcidlink{0000-0002-5436-9845} \and
            Jan Cami \inst{\uwo, \wspace, \seti} \orcidlink{0000-0002-2666-9234} \and
            Am\'elie Canin \inst{\toulouse} \orcidlink{0000-0002-7830-6363} \and
            Ryan Chown \inst{\uwo,\wspace,\osu} \orcidlink{0000-0001-8241-7704} \and
            Emmanuel Dartois \inst{\paris} \orcidlink{0000-0003-1197-7143} 
            Asunción Fuente \inst{\csic} \orcidlink{0000-0001-6317-6343} \and
            Javier R. Goicoechea \inst{\csic} \orcidlink{0000-0001-7046-4319} \and
            Emilie Habart \inst{\paris} \orcidlink{0000-0001-9136-8043} \and
            Olga Kannavou \inst{\paris} \and
            Baria Khan \inst{\uwo} \orcidlink{0000-0002-4350-3017} \and
            Thomas S.-Y. Lai \inst{\ipac} \orcidlink{0000-0001-8490-6632} \and
            Takashi Onaka \inst{\tokyoastro} \orcidlink{0000-0002-8234-6747} \and
            Dries Van De Putte \inst{\uwo, \wspace} \orcidlink{0000-0002-5895-8268} \and
            Ilane Schroetter \inst{\toulouse} \orcidlink{0000-0002-1099-7401} \and
            Ameek Sidhu \inst{\uwo} \and
            Alexander G.~G.~M. Tielens \inst{\maryland} \orcidlink{0000-0003-0306-0028} \and
            Boris Trahin \inst{\paris, \stsci} \orcidlink{0000-0001-5875-5340} \and
            Yong Zhang \inst{\sunyatsen} \orcidlink{0000-0002-1086-7922}
          }
          
   \institute{
   \amesname; \email{alexandros.maragkoudakis@nasa.gov} \and
   \uwoname \and
   \wspacename \and
   \setiname \and
   \umichname \and
   \toulousename \and
   \uohname \and
   \osuname \and
   \parisname \and
   \csicname \and
   \ipacname \and
   \tokyoastroname \and
   \marylandname \and
   \stsciname \and
   \sunyatsenname
             }
 
  \abstract
   {JWST observations of the Orion Bar have revealed rich and diverse Polycyclic Aromatic Hydrocarbon (PAH) emission. These observations allow for the first time a comprehensive characterisation of the charge state and size of the PAH population on morphologically resolved PDR scales, properties closely linked to physical conditions of their inhabiting environments.}
   {We investigate the evolution of the PAH population's charge state and size across key physical zones in the Orion Bar, which include the \HII{} region, the atomic PDR (APDR), and three bright H$\,\textsc{i}$/H$_{2}$ dissociation fronts (DF1, DF2, and DF3). We connect  changes in PAH charge and size as probed by empirical  emission proxies with the varying physical properties of their surrounding environments.}
   {Utilising the NASA Ames PAH Infrared Spectroscopic Database (PAHdb) and the pyPAHdb spectral modelling tool, we analysed the MIRI-MRS observations of the Orion Bar from the "PDRs4All" JWST Early Release Science Program. Decomposition and modelling was performed on the $5-15$ \micron{} spectrum across the entire JWST mosaic, as well as on the weighted average spectra of the five key physical zones.}
   {pyPAHdb modelling reveals the fractional contribution of the different PAH charge states and sizes to the total PAH emission across the Orion Bar. Cationic PAH emission peaks in the APDR region, where neutral PAHs have minimal contribution. Emission from neutral PAHs peaks in the \HII{} region that consists of emission from a face-on PDR associated to the background OMC-1 molecular cloud, and in the molecular cloud regions past DF2. PAH anions are observed deep within the DF2 and DF3 zones. Small and medium sized PAHs make up $\sim 70\%$ of the PAH emission across the mosaic, with the peak of the small PAH emission found between the DF2 and DF3 zones. The average PAH size in the Orion Bar ranges between $\sim 60-74$ \Nc{}. The modelling reveals regions of top-down PAH formation at the ionisation front, and bottom-up PAH formation within the molecular cloud region. The PAH ionisation parameter $\gamma$ ranges between $\sim 2-9\times10^4$. Intensity ratios which are empirical tracers of PAH ionisation ($I_{6.2}/I_{11.2}$, $I_{7.7}/I_{11.2}$, $I_{8.6}/I_{11.2}$) scale well with $\gamma$ in regions encompassing edge-on or face-on PDR emission, but their correlation weakens within the molecular cloud zone.}
   {Modelling of the $5-15$ \micron{} PAH spectrum with pyPAHdb achieves comprehensive characterization of the net contribution of neutral and cationic PAHs across different environments, whereas empirical PAH proxy intensity ratio tracers can be highly variable and unreliable outside regions dominated by PDR emission. The derived average PAH size in the different physical zones is consistent with a view of PAHs being more extensively subjected to ultraviolet processing closer to the ionisation front, and less affected within the molecular cloud.}


\authorrunning{Maragkoudakis et al.}
\titlerunning{PDRs4All XVIII: PAH Ionisation and Size Across the Orion Bar}
\maketitle
%

\section{Introduction}

A set of prominent emission features at 3.3, 6.2, 7.7, 8.6, 11.2, 12.7, 16.4, and 17.4 \micron{}, with weaker features at surrounding wavelengths, are common to the infrared (IR) spectra of numerous astronomical sources. This emission from the so-called aromatic infrared bands (AIBs) is generally attributed to the family of polycyclic aromatic hydrocarbons (PAHs) and are due to vibrational emission of the carbon–carbon and carbon–hydrogen (C–H) bonds in the PAH molecules that become excited upon the absorption of far-ultraviolet (FUV; 6--13.6 eV) photons \citep[][]{Leger1984, Allamandola1985}. PAH emission has been detected in the spectra of Milky Way sources, including planetary and reflection nebulae \citep[e.g.,][]{Bregman1989, Beintema1996, Peeters2002a, Werner2004}, protoplanetary disks \citep[e.g.,][]{Vicente2013}, \HII{} regions \citep[e.g.,][]{Bregman1989, Peeters2002b}, and in a wide variety of galaxies and galactic nuclei \citep[e.g.,][]{Hony2001, Peeters2004, Smith07b, Galliano2008, Sandstrom2012, Boersma2018, Maragkoudakis2018a, Zang2022, Lai2022, Lai2023, Garcia-Bernete2022, Maragkoudakis2022, Chastenet2023, Egorov2023, Rigopoulou2024, Maragkoudakis2025}.

PAH emission can offer unique insights into the local physical conditions of their host environments \citep[e.g.,][]{Galliano2008, Boersma2015, Pilleri2015, Stock2017, Peeters2017, Maragkoudakis2022}. For instance, the intensity of the UV radiation field, $G_0$, in terms of the Habing field \citep[][]{Habing1968} can be assessed through the empirical calibration of the $I_{6.2}/I_{11.2}$ PAH band ratio and the ionisation parameter $\gamma \propto G_0 T^{1/2}/n_e$, where \textit{T} is the gas temperature and $n_e$ is the electron density \citep[e.g.,][]{Galliano2008, Maragkoudakis2022}. The unparalleled spatial resolution and sensitivity of JWST now allow calibration of these diagnostic tracers on spatially resolved scales within astronomical sources.

The study of PAH emission in photo-dissociation regions (PDRs) is of particular interest, as it encompasses emission from zones with different physical conditions and chemical stratification. PDRs are the interfaces between molecular gas and the surrounding galactic medium, illuminated by far-ultraviolet (FUV) photons (6 eV $< E <$ 13.6 eV) from young massive stars. The Orion Bar is a prototypical strongly UV-irradiated PDR in the Orion Nebula (M42). Due to its nearly edge-on view and very high surface brightness, the Orion Bar is one of the most extensively studied sources \citep[][]{Tielens1993, Hogerheijde1995, Kassis2006, Goicoechea2016} and an excellent candidate to examine the evolution of PAH emission, the characteristics of its carriers, and their connection to the local physical conditions across the PDR. 

In this work we characterize the fractional contribution of the different PAH charge states and sizes to the total PAH emission by performing modelling to the $5-15$ \micron{} PAH emission spectrum across the Orion Bar's JWST MIRI-MRS mosaic. This paper is structured as follows. Section \ref{sec:obs-analysis} describes the observations, data reduction, and measurement of the PAH emission bands. Section \ref{sec:results} presents the analysis and results. A discussion and summary of our conclusions are given in Sections \ref{sec:discussion} and \ref{sec:Summary} respectively.

\section{Observations and analysis} \label{sec:obs-analysis}

\subsection{Anatomy of the Orion Bar}

PDRs are mostly neutral regions in the interstellar medium (ISM) in which the physics and chemistry are regulated by FUV photons from nearby massive stars. Four characteristic zones are typically identified across PDRs: the ionisation front (IF) where the gas converts from fully ionised to fully neutral, the atomic PDR (APDR) where hydrogen exists primarily in atomic form, the H$_2$ dissociation front (DF) where molecular hydrogen dissociates due to FUV radiation, and the molecular zone where the gas is nearly fully molecular. 

The nearly edge-on geometry of Orion Bar's PDR permits spatially resolved observations of the different (neutral atomic, and molecular) layers across the IF and DFs \citep[see][their Fig. 14]{Peeters2024}. Strong UV radiation from $\theta^{1}$ Ori C, the most massive and brightest star in the Trapezium cluster, powers and shapes the \HII{} region in the Orion Nebula. The IF, located at a physical distance of 0.27 pc from $\theta^{1}$ Ori C, separates the edge of the PDR from the surrounding \HII{} region. Beyond the IF, the first layers of the PDR are predominantly neutral and atomic, the so-called `atomic PDR' (APDR) zone. At $\sim 0.03$ pc from the IF the FUV photon flux is sufficiently attenuated so that most of the hydrogen becomes molecular, marking the beginning of the molecular PDR. Three such successive H$_2$ ridges representing three edge-on dissociation fronts (DF1, DF2, DF3) are located at increasing distances from the IF. We note that emission from PAHs and other PDR tracers (H$_2$, CO, CH$^+$, [OI] 63 and 145 \micron{}, and [CII] 158 \micron{}) in the \HII{} region, originates from the background face-on PDR in OMC-1 \citep[e.g.][]{Bernard-Salas2012, Parikka2018, Goicoechea2019, Knight2022, Habart2024, Peeters2024}. Beyond DF3 (BDF3), the emission suggests another layer of foreground or background PDR emission. Detection of [CI] 609 \micron{} emission component at velocities different from the Bar \citep[see Fig. 5 of][]{Goicoechea2025} indicates a translucent component in the foreground / background of the Bar. However, \cite{Goicoechea2025} did not detect significant C$_2$H hydrocarbon emission from the background OMC-1 gas. Hence, additional observations are required to characterize the nature of this region. Two proto-planetary disks are also present within the APDR zone, the proplyds 203–504 and 203-506 \citep[][]{Bally2000, Berne2023, Berne2024}, although not embedded in the PDR  rather located in the line of sight toward the Bar \citep[e.g.,][]{Haworth2023}. Here, we focus on the characterization of the PAH emission across the different zones in the Orion's Bar PDR, i.e., the \HII{} region, the APDR, and the three dissociation fronts and their surroundings.

\subsection{Observations and data reduction}
The JWST Early Release Science (ERS) program PDRs4All\footnote{\url{https://pdrs4all.org}} \citep[\textit{Radiative feedback from massive stars}, ID1288;][]{Berne2022b}, has obtained NIRCam and MIRI imaging \citep[][]{Habart2024} as well as NIRSpec and MIRI-MRS spectral mapping mode observations \citep{Peeters2024, Chown2024, VanDePutte2024} of the Orion Nebula. The integral field unit (IFU) observations cover a $9 \times 1$ tiles mosaic, positioned to achieve overlap between the NIRSpec and MIRI-MRS channel 1 field-of-views (FOV). 

From the assembled spectral cubes, a set of five template spectra have been extracted from apertures selected to be representative of the five key physical zones of the Orion Bar: the \HII{} region, the atomic PDR (APDR), and the three bright H$\,\textsc{i}$/H$_{2}$ dissociation fronts (DF1, DF2, and DF3) corresponding to three molecular hydrogen (H$_{2}$) filaments, as identified in the NIRSpec FOV \citep[][]{Peeters2024}.

In this work, we focus on the MIRI-MRS observations, utilising the first three MRS channels and all three sub-bands within each channel (short, medium, and long), which provide coverage of the main PAH bands, i.e. at 6.2, 7.7, 8.6, and 11.2 \micron{}. We perform spectral decomposition and subsequent modelling of the PAH emission spectrum across the entire mosaic, as well as on the weighted average (template) spectra of the five key physical zones. The PAH emission spectrum modelling is performed using the NASA Ames PAH Infrared Spectroscopic Database\footnote{\url{https://www.astrochemistry.org/pahdb}} \citep[PAHdb;][Ricca et al. in prep.]{Bauschlicher2010, Boersma2014a, Bauschlicher2018, Mattioda2020}, which consists of the largest collection of quantum-chemically computed absorption spectra of PAHs of various structures, charge states, sizes, and compositions, accompanied with a suite of modelling tools\footnote{\url{https://pahdb.github.io}}. 

\subsection{Spectral decomposition}

\begin{figure*}
    \centering
    \includegraphics[scale=0.40]{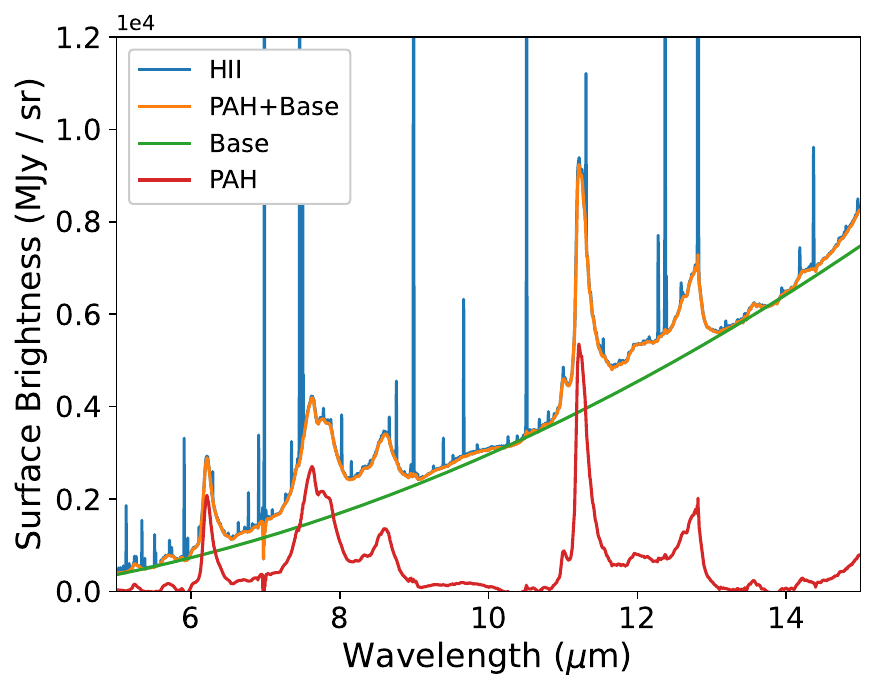}
    \includegraphics[scale=0.40]{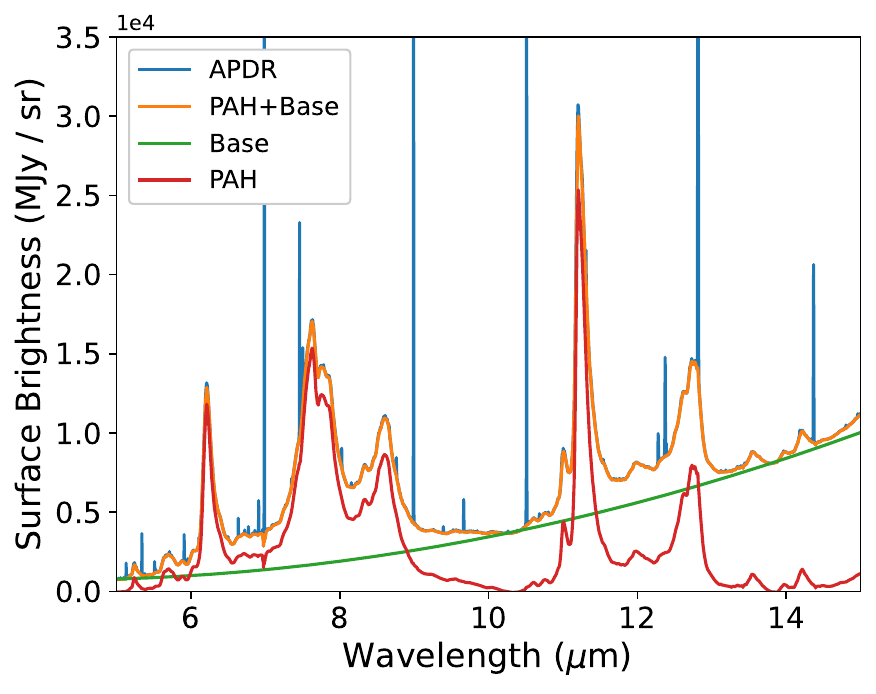}
    \includegraphics[scale=0.40]{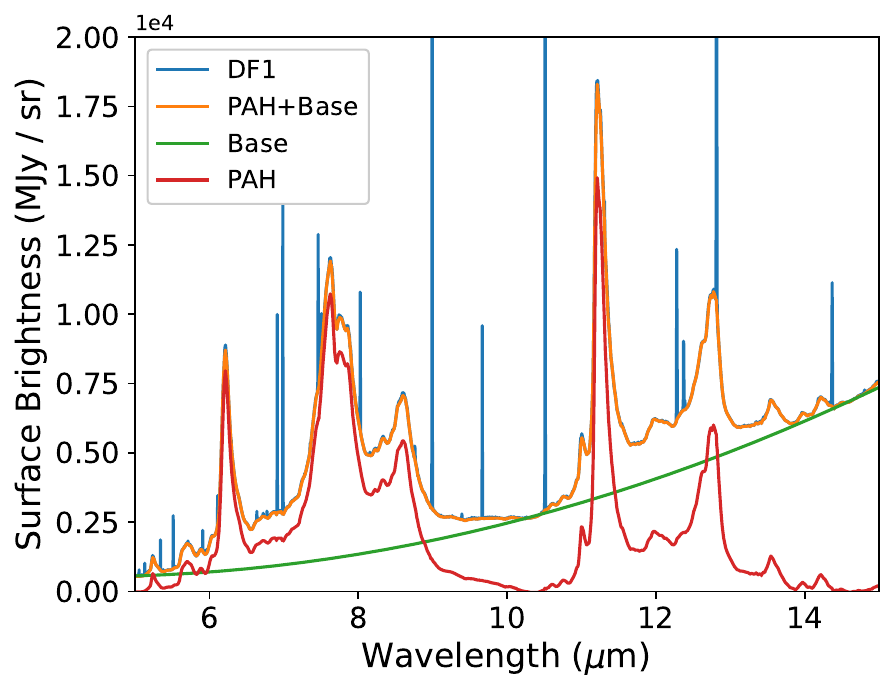} \\
    \includegraphics[scale=0.40]{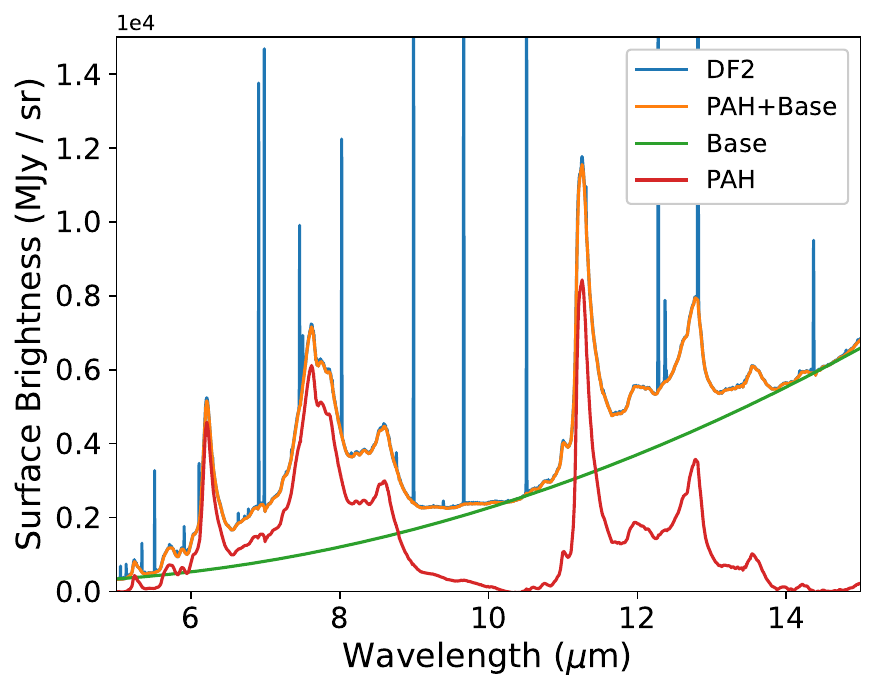}
    \includegraphics[scale=0.40]{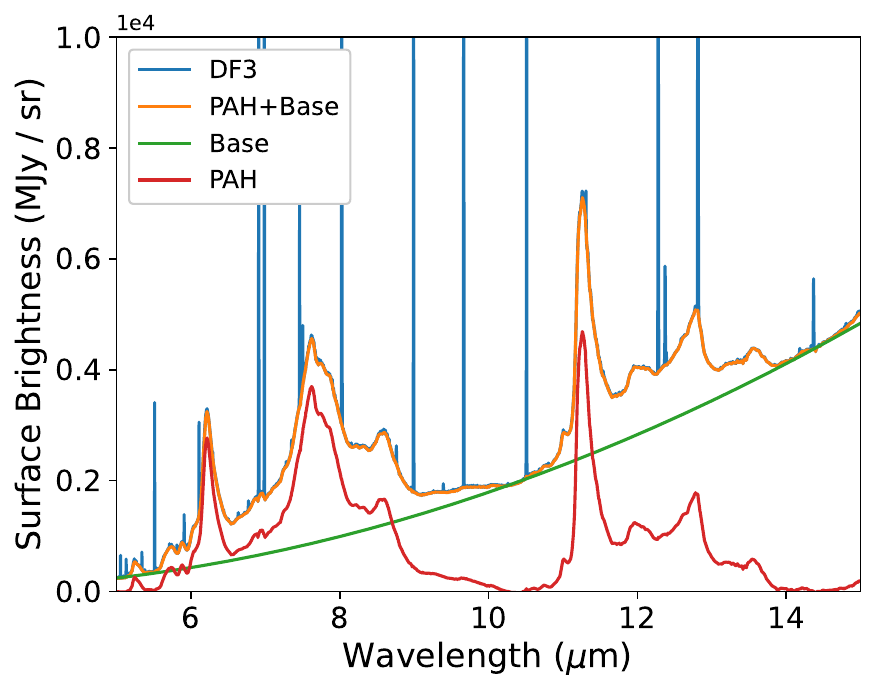}
    \caption{Spectral decomposition of the five Orion Bar template spectra in the 5--15 \micron{} wavelength region. The template spectra are shown in blue, the spectrum after the emission lines removal is shown in orange, the fitted baseline in green, and the resulting isolated PAH emission spectrum in red.} 
    \label{fig:template_decomp}
\end{figure*}

Isolation of the PAH emission component requires removal of the emission lines present in the spectrum (e.g. H\,$_{\textrm{I}}$ recombination lines, ro-vibrational and pure rotational H$_2$ lines, forbidden atomic lines, etc., (see \citealt{Peeters2024} and \citealt{VanDePutte2025}) and the underlying continuum. We achieve this in two steps using the \textsc{pybaseline}\footnote{\url{https://pybaselines.readthedocs.io}} library of algorithms. In the first step, the \textsc{morphological} algorithm is used to create the modelling curve.  This performs a morphological opening transformation on the data and then selects the element-wise minimum between the opening and the average of a morphological erosion and dilation of the opening. The "half-window" parameter defines the size of the window used for the morphological operators, allowing capture of only broad emission features, i.e. the PAH bands, excluding all the narrow emission lines in the spectrum (Figure \ref{fig:template_decomp}, orange lines). In the second step, a baseline is fitted and subtracted from the resulting spectrum in step one, using the \textsc{imodpoly}  (Improved Modified Polynomial) algorithm (Figure \ref{fig:template_decomp}, green lines), which uses thresholding to iteratively fit a polynomial baseline to data. After the baseline subtraction the pure PAH emission spectrum is obtained (Figure \ref{fig:template_decomp}, red lines). Additionally, this process allows one to obtain the emission line spectrum by subtracting both the baseline and the PAH spectrum from the input observational spectrum. We note that no correction for extinction is applied to the spectra. Because the Orion Bar spectra do not exhibit strong absorption features (e.g., silicate absorption or ice features), \textsc{pybaseline} is well-suited for extracting the underlying continuum. The spectral decomposition is performed on the entire MIRI-MRS mosaic and the five template spectra.

\subsection{pyPAHdb modelling}

The modelling of the PAH emission spectra is performed with the pyPAHdb\footnote{\url{https://github.com/PAHdb/pyPAHdb}} tool. pyPAHdb was developed as one of the science-enabling products for the PDRs4All program, and is a streamlined version of the broader  AmesPAHdbPythonSuite\footnote{\url{https://github.com/PAHdb/AmesPAHdbPythonSuite}} package from the PAHdb suite of tools. A first description of pyPAHdb was given in \cite{Shannon2018}. pyPAHdb is designed to quickly and conveniently fit the PAH emission component of a (JWST) spectrum and break it down in terms of PAH charge and size, whereas the AmesPAHdbPythonSuite allows more flexibility on the modelling parameters and the selection of the pool of PAHs. pyPAHdb streamlines the database-fitting by using a \emph{pre-computed} matrix of highly over-sampled synthesized PAH emission spectra, under a specific modelling configuration. In the latest version of pyPAHdb the matrix is built from version 4.00-$\alpha$ of the library of computed spectra \citep[][]{Maragkoudakis2025}, using pure PAH molecules of different charge states (neutral, cation, and anion PAHs), and sizes ($20 <$ \Nc{} $< 384$). The emission spectrum is calculated utilising the `cascade' emission model using an excitation energy of 7 eV, and each transition is convolved with a Gaussian emission profile with a FWHM of 15.0~cm$^{\rm -1}$. \footnote{The effects of different modelling configurations and PAHdb library versions on the results have been extensively examined in \cite{Maragkoudakis2025}.} Given that PAHs absorb an average photon energy of 8.1 eV at the IF \citep{Knight2021} and the average absorbed photon energy decreases as the radiation field is attenuated as we progress into the PDR, we adopt an average photon energy of 7 eV for our analysis.

pyPAHdb returns the breakdown of the contributing PAHs in the fit in terms of the PAH charge and size. Specifically, the PAH charge breakdown provides the fractional contribution of the neutral, cationic, and anionic PAHs to the total PAH emission, while the PAH size breakdown provides the fractional contribution of the small (\Nc{} $\leq 50$), medium ($50 <$ \Nc{} $\leq 70$), and large (\Nc{} $> 70$) PAHs in the fit. pyPAHdb has been especially geared to work with spectral mosaics and, as such, it also delivers maps of the cation-to-neutral fractions and the weighted average \Nc{} from the fitting weights of the contributing PAHs in the fit. 

\begin{figure*}
    \centering
    \includegraphics[scale=0.52]{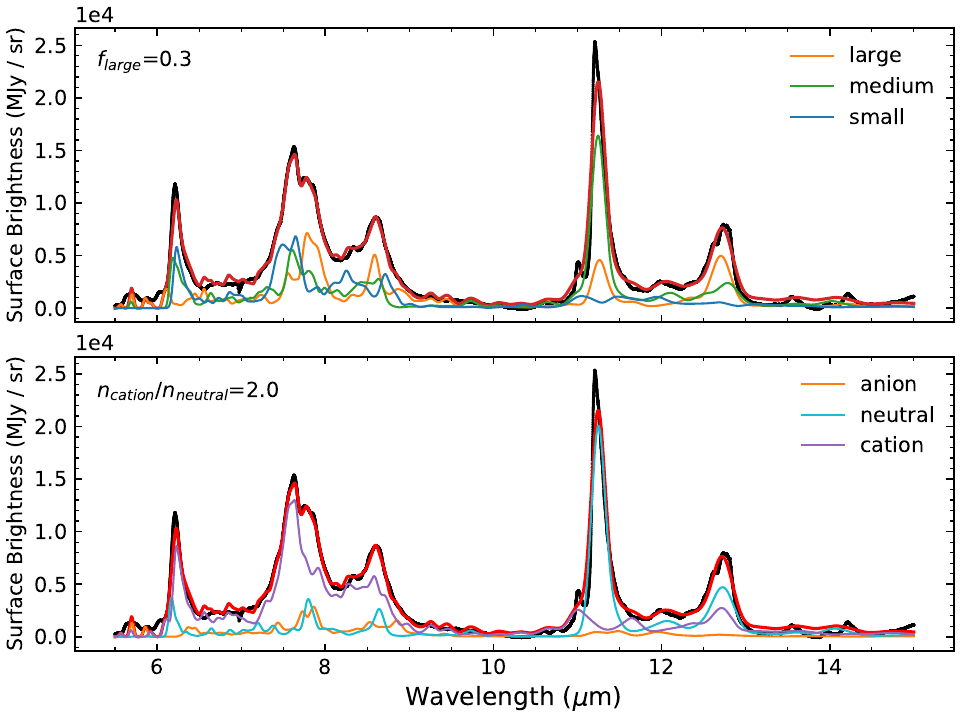}
    \includegraphics[scale=0.52]{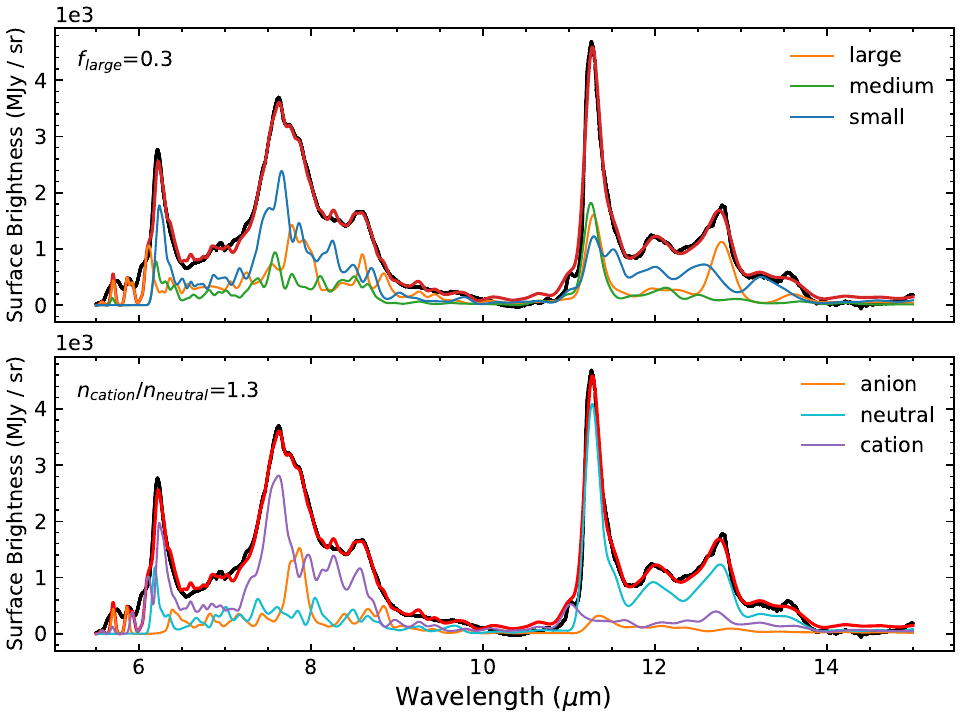}
    \caption{PAH size (top sub-panels) and charge breakdown (bottom sub-panels) for the spectral templates of the APDR (left) and DF3 (right). Observations are shown in black, total fit in red, and the separate breakdown components are given in the legend of each panel.}
    \label{fig:pyPAHdb_templates}
\end{figure*}

\begin{figure}
    \centering
    \includegraphics[scale=0.75]{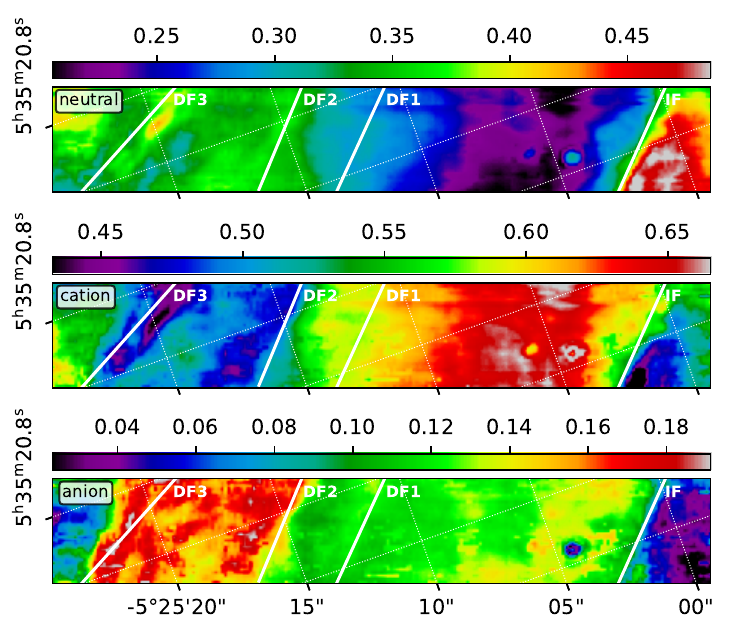}
    \caption{PAH charge breakdown maps. The map of neutral PAHs is shown in the top panel, the PAH cations map in the middle panel, and the PAH anions map in the bottom panel. The colorbar shows the respective charge fractions, and, for each map, the range is scaled between 0.5\% and 99.5\% percentile. The colourmap has been chosen to emphasize structure.}
    \label{fig:charge_breakdown_maps}
\end{figure}

\begin{figure}
    \centering
    \includegraphics[scale=0.75]{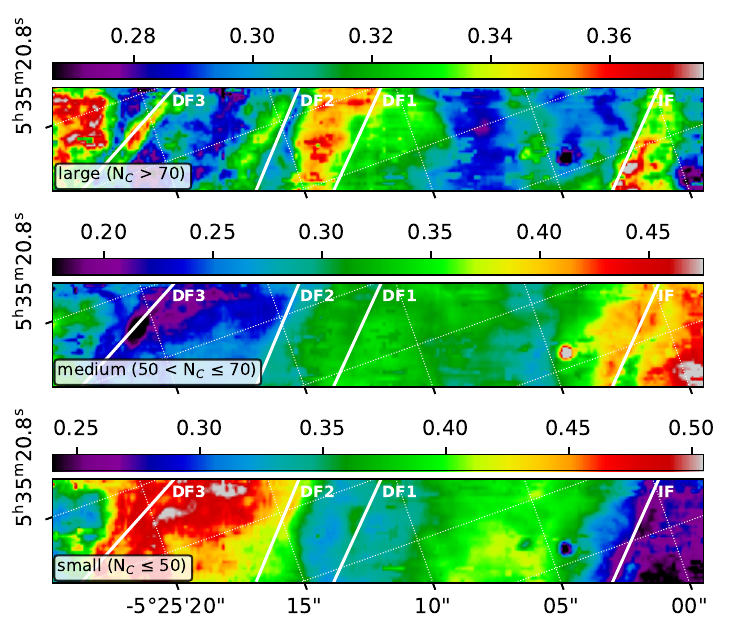}
    \caption{PAH size breakdown maps. The map of large PAHs (\Nc{} $> 70$) is shown in the top panel, the medium-sized PAHs ($50 <$ \Nc{} $\leq 70$) map in the middle panel, and the small PAHs (\Nc{} $\leq 50$) in the bottom panel. The colour bar shows the respective size fractions, and for each map, the range is scaled between 0.5\% and 99.5\% percentile.}
    \label{fig:size_breakdown_maps}
\end{figure}

Modelling with pyPAHdb was performed on both the entire MIRI-MRS mosaic and the template spectra of the five key physical zones of the PDR. Figure \ref{fig:pyPAHdb_templates} shows an example of the PAH charge and size breakdown of the APDR and DF3 template regions, and plots of the remaining regions are presented in Appendix \ref{sec:template_plots} (Figure \ref{fig:appendix_pyPAHdb_templates}). The figure demonstrates the high level of detail the modelling is able to reproduce \citep[see also][]{Maragkoudakis2025}, and reiterates the effect of charge on a PAH's spectrum where the bulk of the intensity shifts from the 11-14 to the 6-9 \micron{} region upon ionization \citep[][]{Allamandola1999}. Figures \ref{fig:charge_breakdown_maps} and \ref{fig:size_breakdown_maps} present the different maps of the PAH charge breakdown and the PAH size, respectively. The maps of the PAH ionisation parameter ($\gamma$; see Section \ref{sec:gamma}) and weighted average \Nc{} are shown in Figures \ref{fig:gamma_map} and \ref{fig:nc_map}, respectively. The map of the modelling error ($\sigma_{pyPAHdb}$) is presented in Appendix \ref{sec:pypahdb_unc} (Figure \ref{fig:pyPAHdb_errormap}), which is typically between $10-15\%$. All results need to be considered within the limits of the employed (pyPAHdb) modelling and its sensitivity to parameter choices. Such aspects (e.g., different excitation energies, emission models, line profiles, PAHdb library versions) have been previously thoroughly explored in, for example, \cite{Andrews2015} and \cite{Maragkoudakis2025} and provide confidence in the overall approach.

\subsection{PAH band measurements} \label{sec:pah_band_measurements}

Measurement of the PAH band strengths was performed on the pure PAH emission spectra by integrating the flux within defined wavelength intervals. The wavelength regions were determined by visual inspection after overplotting the PAH emission spectra of the five key physical regions (Figure \ref{fig:pah_bandpasses}). This method, while not necessarily as precise as a multi-component decomposition \citep[e.g.,][]{VanDePutte2025, Khan2025a}, has the advantage of capturing the bulk of the PAH emission, providing a good estimate of  the PAH band intensity ratios efficiently for the $\sim 8,500$ pixels that span the entire mosaic. Figure \ref{fig:pah_bands_maps} presents the PAH intensity maps of the $I_{6.2}$, $I_{7.7}$, $I_{8.6}$, and $I_{11.2}$ \micron{} bands, and Figure \ref{fig:pah_ratios_maps} the maps of the $I_{6.2}/I_{11.2}$, $I_{7.7}/I_{11.2}$, and $I_{8.6}/I_{11.2}$, PAH band intensity ratios. We have also obtained the $I_{3.3}$ and $I_{3.3}/I_{11.2}$ maps from Schefter et al. (in prep.).

\section{Results} \label{sec:results}

\subsection{PAH charge and size across the Orion Bar} \label{sec:charge-size_breakdown}

Figure \ref{fig:charge_breakdown_maps} presents the maps of the PAH charge breakdown, i.e. the fractional contribution of the neutral ($f_{neut}$), cation ($f_{cat}$), and anion ($f_{an})$ PAHs to the total PAH emission. It is clear from Figure \ref{fig:charge_breakdown_maps} that the dominant charge state changes across the FOV, with neutral PAHs dominating in the line-of-sight toward the \HII{} region, cationic PAHs in the atomic PDR, and anionic PAHs in the molecular PDR. The PAH emission in the line of sight towards the \HII{} region is actually originating from the background face-on PDR in OMC-1 \citep[e.g.][]{Peeters2024}. In this region, approximately $40\%-55\%$ of the total PAH emission is due to neutral PAHs, $\sim45\%-60\%$ due to PAH cations, and $\sim5\%$ due to PAH anions. PAH emission in the APDR is dominated by cations ($60\%-70\%$), followed by neutrals (25\%), and then anions ($10\%-15\%$). Moving towards the DF1, and into the DF2 and DF3 zones, where the flux of FUV photons is significantly attenuated \citep[][]{Peeters2024, Habart2024}, $f_{neut}$ increases progressively from 25\% to 40\% while $f_{cat}$ decreases from 60\% to 45\%, and $f_{an}$ has its maximum contribution (up to 20\%) in the DF3 region. In the region beyond DF3, $f_{neut}$ is $\sim30\%-40\%$, $f_{cat}$ is $\sim50\%-65\% $ and $f_{an}$ reaches up to $\sim 20\%$. The peak of PAH anion emission is observed deep within the DF2/DF3 zones. As the FUV radiation is more attenuated deeper into the PDR, a change in the charge balance is expected, such as an increase in neutral and anionic PAHs \citep[][]{BakesTielens1994}, although anions are expected to be present deeper within the molecular cloud.

The PAH size breakdown maps (Figure \ref{fig:size_breakdown_maps}) present the fractional contribution of small ($f_{small}$), medium ($f_{med}$), and large ($f_{large}$) PAHs to the total PAH emission, revealing regions where the different PAH size classes have a prevalent contribution. Large PAHs (\Nc{} $> 70$) contribute overall between $\sim 20\% - 40\%$, with the highest $f_{large}$ found in the \HII{} region close to the IF, between DF1 and DF3, and beyond DF3. Medium sized PAHs ($50 <$ \Nc{} $\leq 70$) contribute $\sim 10\% - 50\%$ to the total PAH emission. 40\% - 50\% of $f_{med}$ is found in the \HII{} region and the APDR region closest to the IF, while the lowest contribution ($\sim 20\% - 25\%$) is between DF2 and DF3. The smaller PAHs (\Nc{} $\leq 50$) have a $\sim 25\% - 50\%$ contribution to the total PAH emission. The highest $f_{small}$ (45\% - 50\%) is observed between DF2 and DF3, deep within the molecular cloud where small PAHs are less prone to photodissociation, and the lowest contribution (25\% - 30\%) is seen in the \HII{} region.  

\subsection{The PAH ionisation parameter ($\gamma$)} \label{sec:gamma}

\begin{figure}
    \centering
    \includegraphics[scale=0.65]{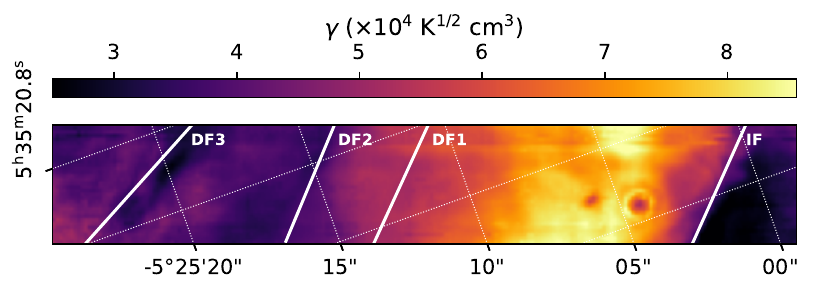} 
    \caption{Map of the PAH ionisation parameter ($\gamma$) obtained from the pyPAHdb modelling. White lines indicate the ionisation and dissociation fronts.}
    \label{fig:gamma_map}
\end{figure}

The ionisation parameter, $\gamma$, provides a description of the PAH ionisation balance as a function of local physical parameters, in particular the intensity of the radiation field ($G_0$), the gas temperature \textit{T$_{gas}$}, and the electron density $n_e$. When assuming two accessible ionisation states and parameters applicable for circumcoronene (C$_{54}$H$_{12}$), which is considered representative of an average interstellar PAH \citep[\Nc{} = 50 - 100;][]{Croiset2016}, the PAH ionisation parameter can be expressed as: 

\begin{equation} \label{eq:npah_ioniz_param}
    \gamma \equiv(G_{0} \times \sqrt{T_{\rm gas}}) / n_{\rm e} = 2.66\, (\frac{n_{\mathrm{PAH^{+}}}}{n_{\mathrm{PAH^{0}}}})\ [\times10^{4}\ \mathrm{K}^{1/2} \mathrm{cm}^{3}],
\end{equation}
where $n_{\mathrm{PAH^{+}}}$ and $n_{\mathrm{PAH^{0}}}$ are the PAH cation and neutral densities, which correspond to $f_{cat}$ and $f_{neut}$ respectively, as determined from the pyPAHdb modelling \citep{Tielens2005, Boersma2014a}.

Figure \ref{fig:gamma_map} presents the map of $\gamma$. Values are ranging between 2- 9 $\times10^{4}\ \mathrm{K}^{1/2} \mathrm{cm}^{3}$. $\gamma$ peaks in the APDR region between the IF and DF1 (6 - 8.5 $\times10^{4}\ \mathrm{K}^{1/2} \mathrm{cm}^{3}$), where $f_{cat}$ is the highest and $f_{neut}$ lowest (Figures \ref{fig:charge_breakdown_maps} - \ref{fig:size_breakdown_maps}). Between DF1 and DF2 $\gamma$ is $\sim 5\times10^{4}\ \mathrm{K}^{1/2} \mathrm{cm}^{3}$, dropping to $\sim 4\times10^{4}\ \mathrm{K}^{1/2} \mathrm{cm}^{3}$ beyond DF2, and has the lowest value ($\sim 3\times10^{4}\ \mathrm{K}^{1/2} \mathrm{cm}^{3}$) in the zone from the IF towards $\theta^{1}$ Ori C, which consists of emission originating from the background face-on PDR.

\subsection{The average PAH size} \label{sec:size}

\begin{figure}
    \centering
    \includegraphics[scale=0.65]{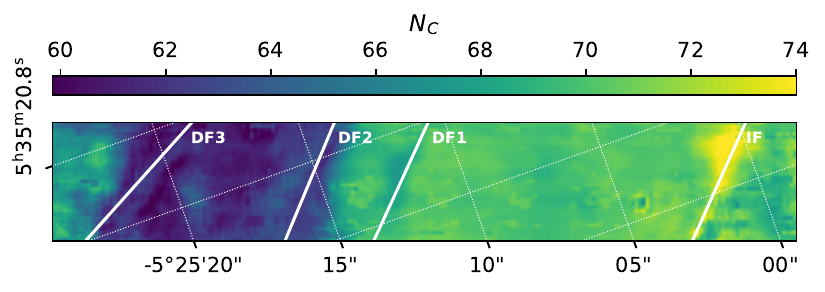}
    \caption{Map of the average \Nc{} obtained from pyPAHdb modelling. White lines indicate the ionisation and dissociation fronts.}
    \label{fig:nc_map}
\end{figure}

Figure \ref{fig:nc_map} presents the map of the average PAH size in terms of \ANc{} weighted by their emission contribution to the total emission. \ANc{} ranges between 60 and 74 C atoms. It peaks at the IF where small PAHs are less present due to photodissociation (see also Figure \ref{fig:size_breakdown_maps}, bottom panel). From the IF and up to DF2 the \ANc{} is $\sim 70$ C atoms, and drops to $\sim 62$ C atoms in the molecular hydrogen region between DF2 and DF3. In the region beyond DF3 and in the \HII{} zone, the \ANc{} is $\sim 66 -70$ C atoms.  

\subsection{Modelling of the five key physical zone spectra} \label{sec:templates}

\begin{table*}[]
    \caption{pyPAHdb derived parameters for the five template spectra (\HII{}, APDR, DF1, DF2, DF3) and average parameter values of the regions between the key physical zones (\HII{}-IF, APDR-DF1, DF1-DF2, DF2-DF3, DF3-BDF3).}
    \centering
    \begin{tabular}{cccccccccc}
    \hline
    \hline
        Zone & $f_{neut}$ & $f_{cat}$ & $f_{an}$ & $f_{sm}$ & $f_{med}$ & $f_{lar}$ & $\gamma$ & \ANc{} \\
    \hline
        \HII{} & 0.46 & 0.51 & 0.03 & 0.25 & 0.43 & 0.32 & 2.9 & 70 \\
        \HII{}-IF & 0.44 & 0.51 & 0.05 & 0.26 & 0.41 & 0.32 & 2.7 & 70 \\
        APDR & 0.29 & 0.60 & 0.11 & 0.30 & 0.39 & 0.31 & 5.5 & 71 \\
        APDR-DF1 & 0.25 & 0.63 & 0.12 & 0.36 & 0.33 & 0.31 & 6.7 & 70 \\
        DF1 & 0.30 & 0.58 & 0.12 & 0.35 & 0.31 & 0.34 & 5.1 & 69 \\
        DF1-DF2 & 0.32 & 0.56 & 0.12 & 0.36 & 0.30 & 0.34 & 4.7 & 67 \\
        DF2 & 0.36 & 0.48 & 0.16 & 0.42 & 0.25 & 0.33 & 3.6 & 63 \\
        DF2-DF3 & 0.35 & 0.49 & 0.16 & 0.46 & 0.24 & 0.30 & 3.7 & 62 \\
        DF3 & 0.36 & 0.47 & 0.17 & 0.47 & 0.20 & 0.33 & 3.5 & 62 \\
        BDF3 & 0.36 & 0.52 & 0.12 & 0.41 & 0.26 & 0.33 & 3.9 & 65 \\
    \hline
    \end{tabular}
    \tablefoot{From left to right: the fractional contribution of neutral, cation, anion, small, medium, and large PAHs in the fit, the ionisation parameter $\gamma$ ($\times 10^4 K^{1/2} cm^3$), and the average \Nc{}. Note that the regions between the key physical zones include emission of the template spectra.}
    \label{tab:templates_breakdown}
\end{table*}

Five key regions have been identified in the NIRSpec \citep[][]{Peeters2024} and NIRCam \citep[][]{Habart2024} observations, representative of the key physical zones of the Orion Bar mosaic: the \HII{} region, the atomic PDR, and three bright H$\,\textsc{i}$/H$_{2}$ dissociation fronts (see Figure \ref{fig:pah_bands_maps}). Here, we perform pyPAHdb modelling of the five MIRI-MRS template spectra \citep[][]{Chown2024}, which are the flux-weighted average of the pixels in the extraction apertures described in \cite{Peeters2024}. Table \ref{tab:templates_breakdown} presents derived parameters for the five template spectra, and parameter values of the regions between the key physical zones denoted by the IF and DFs (\HII{}-IF, APDR-DF1, DF1-DF2, DF2-DF3, DF3-BDF3). The PAH charge and size breakdowns, including the derivative $\gamma$ and \ANc{} parameters, are in agreement with the ranges of the broader and surrounding regions in the mosaic (Sections \ref{sec:charge-size_breakdown} - \ref{sec:size} and Table \ref{tab:templates_breakdown}) in which the five regions reside.

\subsection{PAH band strengths and relative intensities}

\begin{figure}
    \centering
    \includegraphics[scale=0.58]{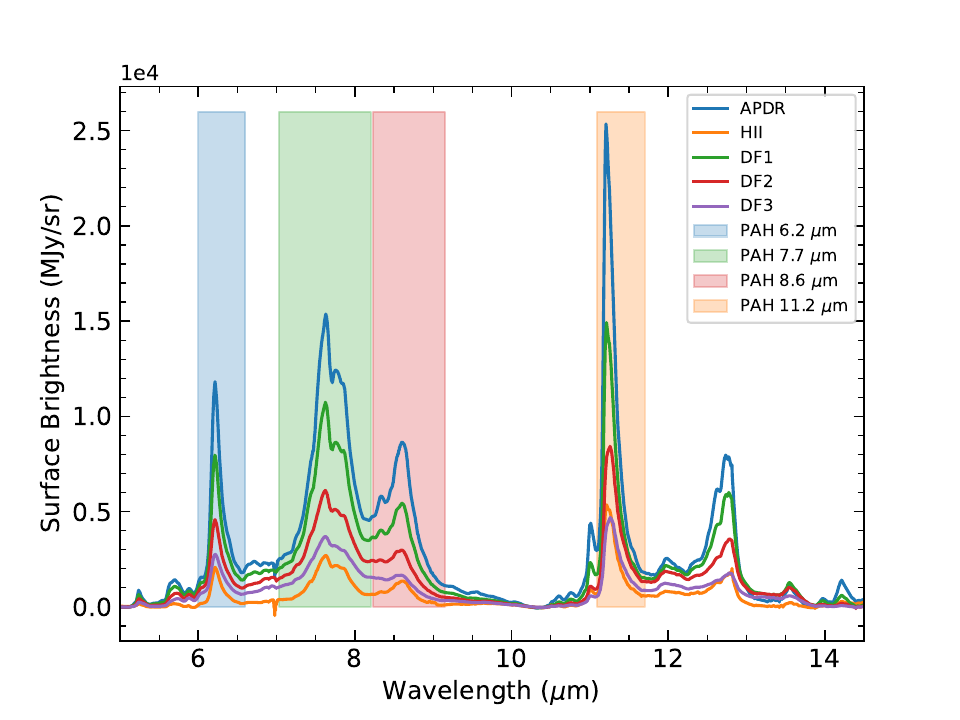}
    \caption{Wavelength regions adopted for the measurement of the different PAH emission intensities, as deduced from the five spectral templates. The wavelength intervals are defined as follows: PAH$_{6.2 \mu m}$: 6.0 \micron{} - 6.6 \micron{}; PAH$_{7.7 \mu m}$: 7.0 \micron{} - 8.2 \micron{}; PAH$_{8.6 \mu m}$: 8.2 \micron{} - 9.15 \micron{}; PAH$_{11.2 \mu m}$: 11.1 \micron{} - 11.7 \micron{}}.
    \label{fig:pah_bandpasses}
\end{figure}

Empirical calibrations involving the intensities of PAH emission bands sensitive to charge and size, have been used to trace the charge state and size distribution of PAHs within and among astronomical sources. Such calibrations can be determined from models with assumptions on the PAH size distribution, the stochastic heating of PAHs, the absorption efficiencies, etc. \citep[e.g.][]{Schutte1993, Draine2007, Galliano2008, Stock2016, Draine2021}, or derived without such assumptions from the ground up, from quantum-chemically computed spectra of PAH populations with various charge states and sizes \citep[e.g.][]{Cami2011, Ricca2012, Boersma2013, Boersma2015, Andrews2015, Boersma2018, Croiset2016, Maragkoudakis2020, Maragkoudakis2023a, Maragkoudakis2023b}.    

\begin{figure*}
    \centering
    \includegraphics[scale=0.65]{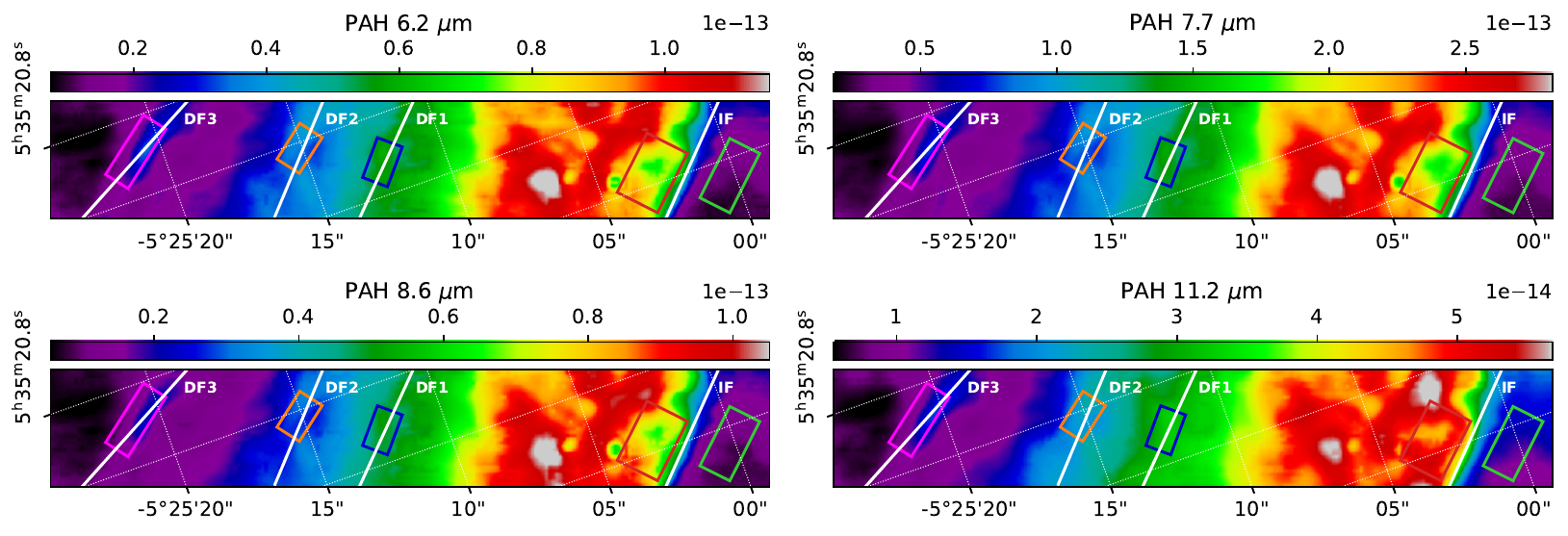} \\
    \caption{The maps of the 6.2 \micron, 7.7 \micron{}, 8.6 \micron{}, and 11.2 \micron{} PAH flux (in erg s$^{-1}$ cm$^{-2}$). The rectangular apertures of the five template regions are indicated, along with lines for the IF and DFs.}
    \label{fig:pah_bands_maps}
\end{figure*}

\begin{figure}
    \includegraphics[scale=0.75]{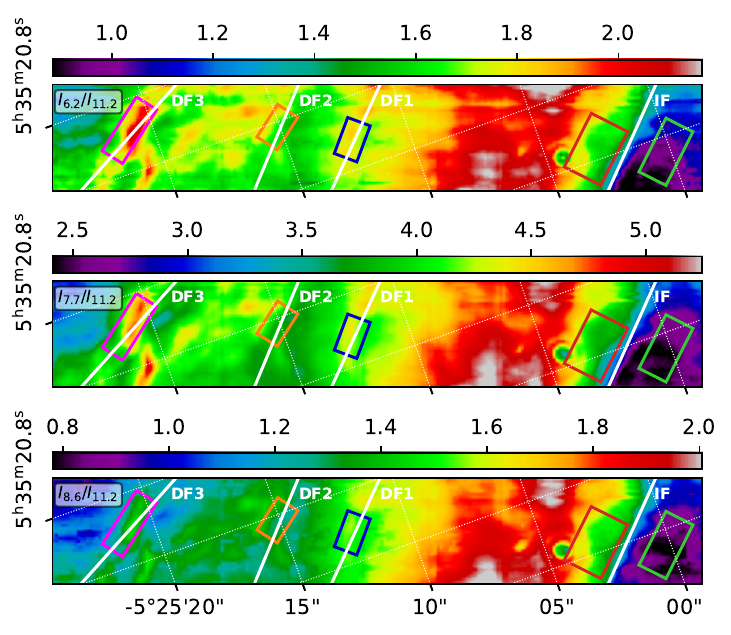}\\
    \caption{PAH band intensity ratio maps. The rectangular apertures of the five template regions are indicated, along lines for the IF and DFs.}
    \label{fig:pah_ratios_maps}
\end{figure}

To examine the dependence of pyPAHdb-derived parameters ($\gamma$, \ANc{}) with the PAH intensity ratios employed in empirical calibrations, we have measured the integrated specific surface brightnesses ("surface brightnesses" hereafter) of the main PAH bands at 6.2, 7.7, 8.6, and 11.2 \micron{} (Section \ref{sec:pah_band_measurements}), and generated their intensity maps, shown in Figure \ref{fig:pah_bands_maps}. We note that the wavelength region used to measure the $I_{8.6}$ may include contribution from weaker features at 8.223 and 8.330 \micron{} \citep{Chown2024}, which can be more prominent in the APDR region (Figure \ref{fig:pah_bandpasses}). These maps were then used to construct $I_{6.2}/I_{11.2}$, $I_{7.7}/I_{11.2}$, and $I_{8.6}/I_{11.2}$ PAH intensity ratio maps (Figure \ref{fig:pah_ratios_maps}). 

The surface brightness of all the PAH bands peaks within the APDR zone \citep[][Schefter et al. in prep., Khan et al. 2025b]{Peeters2024, Habart2024, Pasquini2024, Khan2025a}, between the IF and DF1, gradually decreasing towards the molecular cloud region. Despite this gradual drop in surface brightness, all bands have relatively elevated values on the DF3 boundary compared to its immediate surrounding regions. Regions of relatively weaker emission are found in the \HII{} zone and beyond DF3. The surface brightness in these regions decreases compared to the APDR by a factor of $\sim10$ in the case of the $I_{6.2}$ and $I_{8.6}$ surface brightness, and by a factor of $\sim 5 - 6$ for the $I_{7.7}$ and $I_{11.2}$ surface brightness. Similar trends are observed in the $I_{6.2}/I_{11.2}$, $I_{7.7}/I_{11.2}$, $I_{8.6}/I_{11.2}$ PAH surface brightness ratio maps (Figure \ref{fig:pah_ratios_maps}). The PAH surface brightness ratios peak within the APDR zone, while regions with relatively lower values are in zones beyond DF3 and in the \HII{} region. 

\subsection{$\gamma$ - PAH intensity ratio calibrations}

\begin{figure*}
    \centering
    \includegraphics[scale=0.45]{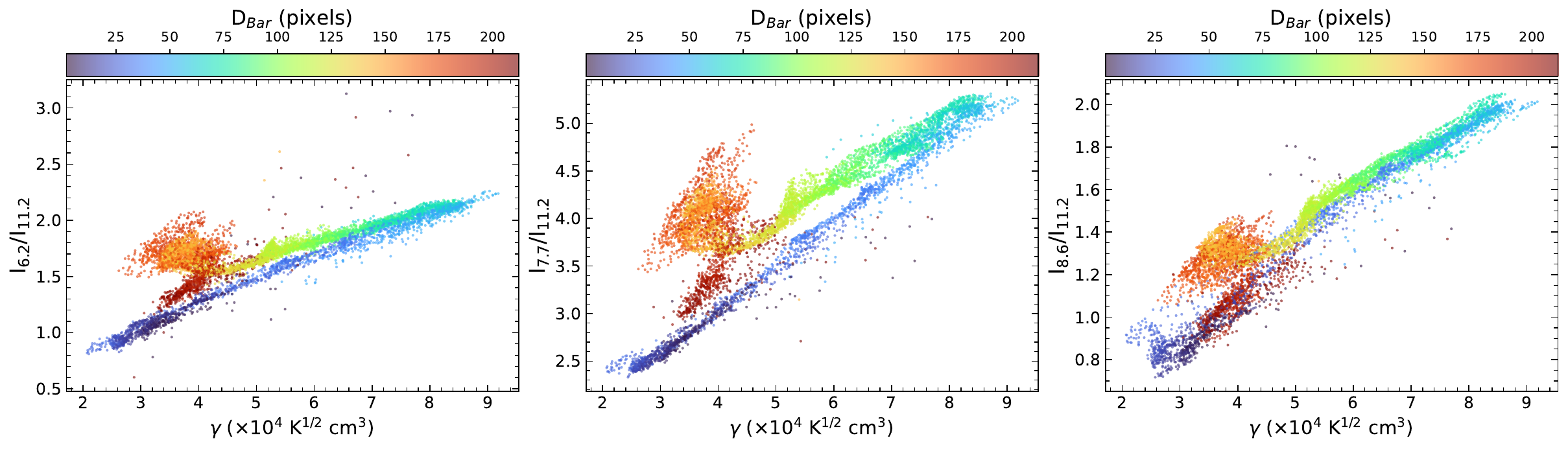}
    \includegraphics[scale=0.45]{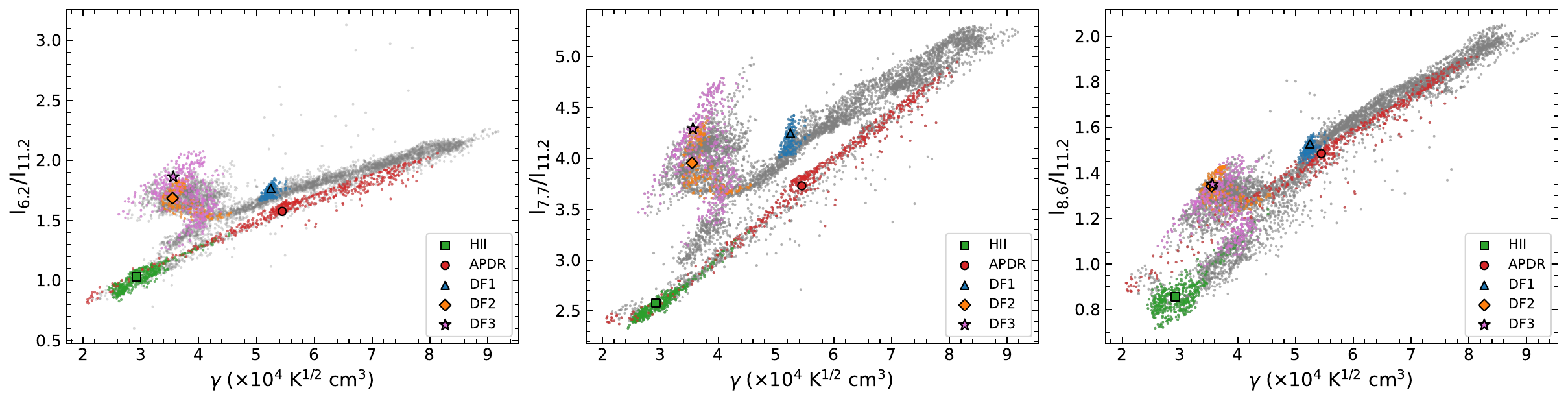}
    \caption{Correlations between the different empirical PAH ionisation proxies ($I_{6.2}$/$I_{11.2}$, left panels; $I_{7.7}$/$I_{11.2}$, middle panels; $I_{8.6}$/$I_{11.2}$, right panels)  and the PAH ionisation parameter ($\gamma$) deduced from pyPAHdb modelling. In the top panels the pixels are colour-coded based on their vertical distance from the edge of the mosaic, which is indicative of the projected distance from $\theta^{1}$ Ori C. In the bottom panels pixels corresponding to each of the five key physical regions are assigned a specific colour (\HII{} in green, APDR in red, DF1 in blue, DF2 in orange, and DF3 in pink), with the average values from each template indicated with a different symbol. Pixels outside the five zones are shown in grey.}
    \label{fig:gamma-pah_ratios}
\end{figure*}

The pyPAHdb-determined ionisation parameter $\gamma$ is compared against empirical tracers of PAH ionisation, i.e., the $I_{6.2}/I_{11.2}$, $I_{7.7}/I_{11.2}$, and $I_{8.6}/I_{11.2}$ relative PAH band intensity ratios (Figure \ref{fig:gamma-pah_ratios}). A linear relation between $\gamma$ and the empirical tracers is observed for the majority of regions (pixels) across the mosaic. Colour-coding of the $\gamma$ - PAH intensity ratio correlation as a function of the distance from the edge of the mosaic, indicative of the distance from $\theta^{1}$ Ori C, reveals a dependence of this correlation with distance, i.e. within the different physical regions. The bottom panels of Figure \ref{fig:gamma-pah_ratios} show the correlation for the individual pixels of each template region. 

The \HII{} and APDR regions span a wide range in $\gamma$ and PAH intensity ratios, with both quantities increasing from the \HII{} zone towards and across the APDR. Both regions present similar $\gamma$-$I_{6.2}/I_{11.2}$ and $\gamma$-$I_{7.7}/I_{11.2}$ correlations ($\gamma$-$I_{6.2,7.7}/I_{11.2}$). Pixels located in the region between the IF and DF1, but outside the APDR template, form a distinctive sequence from the \HII{} and APDR template regions (Figure \ref{fig:gamma-pah_ratios}, bottom panels). Specifically, the $\gamma$-$I_{6.2,7.7}/I_{11.2}$ relations in the regions following the APDR and towards the dissociation fronts, including DF1, have a relatively shallower slope from the corresponding correlations in the APDR--\HII{} regions, with $\gamma$ and the $I_{6.2,7.7}/I_{11.2}$ ratios decreasing with the distance from the Bar. These regions maintain linear correlations among $\gamma$-$I_{6.2, 7.7}/I_{11.2}$, offset from the APDR--\HII{} template sequence, and with a somewhat larger scatter in the $\gamma$-$I_{7.7}/I_{11.2}$ case. The correlations of the regions starting from the edge of the mosaic (towards $\theta^{1}$ Ori C) and up to the DF1 are less separated in the $\gamma$-$I_{8.6}/I_{11.2}$ case, although with more scatter compared to the $\gamma$-$I_{6.2,7.7}/I_{11.2}$ correlations. 

Moving from DF1 towards the DF2 and DF3 regions the correlations between $\gamma$ and the PAH ionisation proxies ($I_{X}/I_{11.2}$) begin to break. However, the pixels in the regions beyond DF3 (dark red pixels in Figure \ref{fig:gamma-pah_ratios}) exhibit a linear $\gamma$-$I_{X}/I_{11.2}$ correlation, albeit with a large scatter. In the $\gamma$-$I_{6.2,7.7}/I_{11.2}$ case, those pixels have a similar correlation with the pixels between the IF and DF1, excluding the APDR template region, while in the  $\gamma$-$I_{8.6}/I_{11.2}$ case they appear in line with the correlation for the \HII{} to DF1 pixels.

\subsection{The \Nc{} - $I_{3.3}/I_{11.2}$ relation}

\begin{figure}
    \centering
    \includegraphics[scale=0.60]{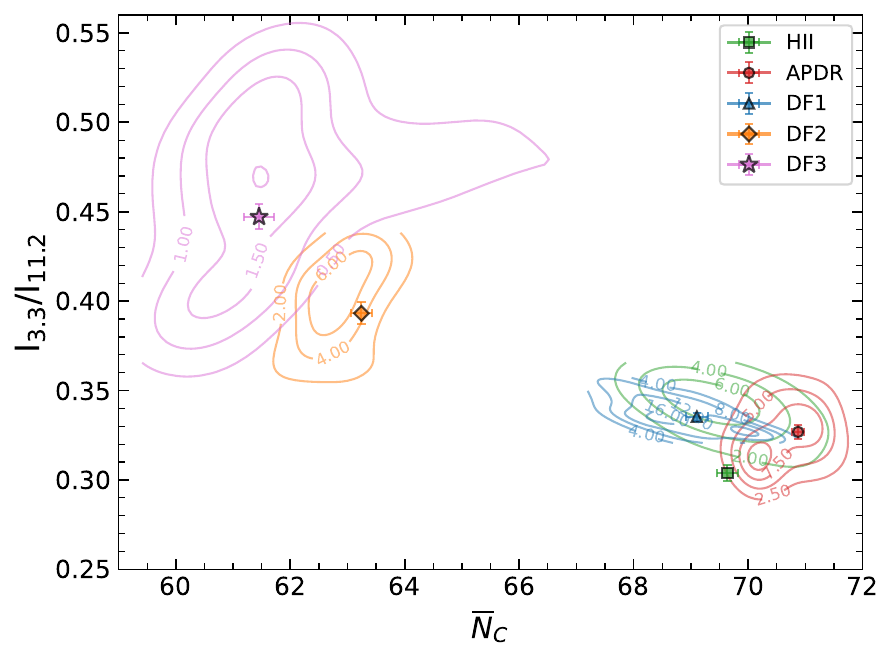}
    \caption{Correlation between the $I_{3.3}$/$I_{11.2}$ PAH size proxy and \ANc{} deduced from pyPAHdb modelling for the five key physical regions. Equidistant contours are drawn up to 3-sigma from the mean value in each template region. Contour labels indicate pixel density, i.e. points per $I_{3.3}$/$I_{11.2}$ per \ANc{}. Region means within 3 standard errors of the mean uncertainty are shown.}
    \label{fig:33_112_vs_Nc_templates}
\end{figure}

The average number of carbon atoms \Nc{} retrieved from the pyPAHdb modelling of the 5--15 \micron{} observations is compared against the well-defined $I_{3.3}/I_{11.2}$ empirical tracer for PAH size \citep[e.g.,][]{Schutte1993, Ricca2012, Croiset2016, Maragkoudakis2020, Knight2021, Maragkoudakis2023a, Maragkoudakis2023b}. For the comparison, we incorporate the PAH 3.3 \micron{} measurements from Schefter et al. (in prep.) and show the $I_{3.3}/I_{11.2}$ -- \ANc{} correlation for the five key physical zones (Figure \ref{fig:33_112_vs_Nc_templates}) along with contours for pixels within 3 standard deviations from the mean in each template region. Moving from the DF3 region towards the APDR, the relative $I_{3.3}/I_{11.2}$ PAH intensity gradually decreases (Schefter et al., in prep.), which follows an increase in \ANc{} (Figure \ref{fig:nc_map}). Towards the IF, smaller PAHs become more prone to photodissociation, resulting in a decrease in the $I_{3.3}$ intensity and an increase in \ANc{}, as intermediate to large PAHs become more abundant (Figure \ref{fig:size_breakdown_maps}). The \HII{} region, where PAH emission originates from the background face-on PDR, diverges from the observed correlation. While the \HII{} region contours of the $3\sigma$ pixels overlap with the DF1 and APDR contours, the average $I_{3.3}/I_{11.2}$ value of the \HII{} template is lower, driven by outliers outside the 3-$\sigma$ cut-off. The unique behaviour of PAH emission in the \HII{} region has been previously noted \citep[][]{Khan2025a} and may be attributed to geometric effects or differences in the characteristics of the UV field at the IF \citep[][]{Peeters2024}.

\begin{figure}
    \centering
    \hspace*{-0.3cm}\includegraphics[scale=0.55]{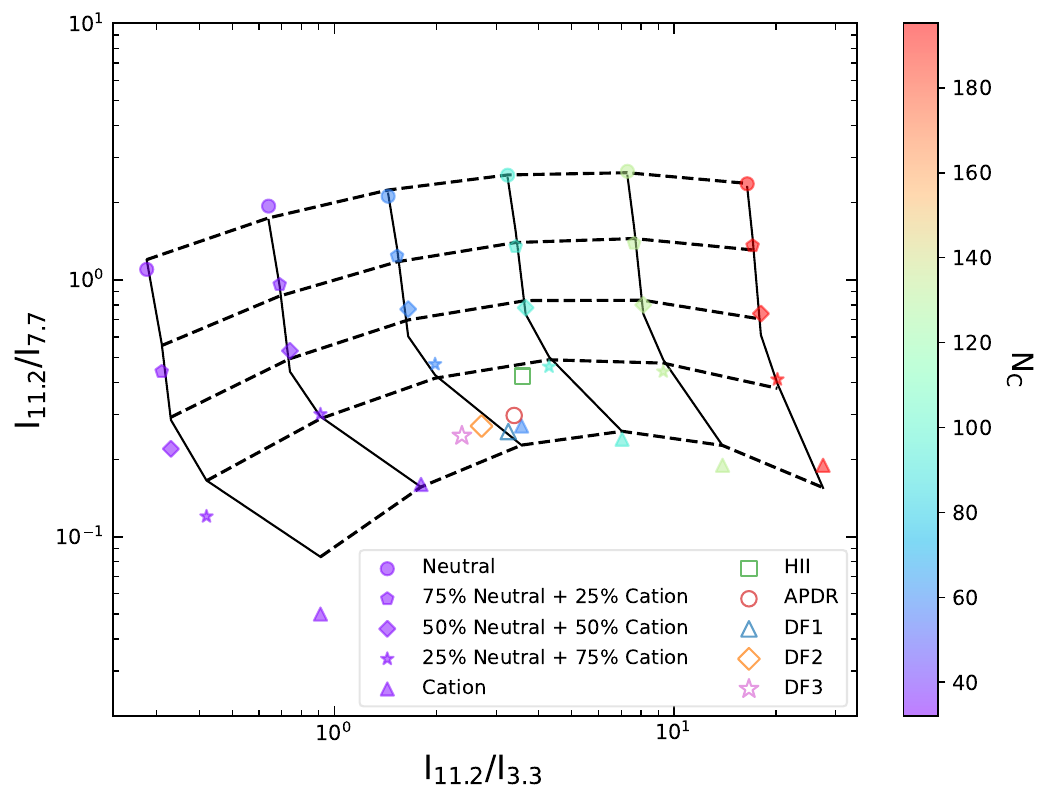}
    \caption{PAH charge--size grid \citep{Maragkoudakis2020} providing an alternative, fitting-independent description to the charge state and \ANc{} of the five key physical regions.}
    \label{fig:charge-size-grid}
\end{figure}

The observed $I_{3.3}/I_{11.2}$ -- \ANc{} correlation is the product of the intrinsic response of different sized PAHs to the UV field, where larger PAHs are contributing most of their emission at longer wavelengths and smaller PAHs are contributing predominantly to the emission at shorter wavelengths \citep[][]{Allamandola1989, Schutte1993}. Therefore, PAH bands with large wavelength separation and of the same charge state (neutral), can effectively trace the different ends of the PAH size distribution, and can therefore describe \ANc{}. The $I_{3.3}/I_{11.2}$ ratio is a particularly good tracker of PAH size because both bands are due to CH stretches and arise primarily from neutral PAHs. Using the \cite{Maragkoudakis2020} PAH charge--size grid we can obtain a fitting-independent description of the size distribution of PAHs in the five template regions (Figure \ref{fig:charge-size-grid}). The \ANc{} of the template regions gradually increases moving from the DF3 to the \HII{} region, as the $I_{11.2}/I_{3.3}$ ratio increases ($I_{3.3}/I_{11.2}$ decreases). The \ANc{} values are calculated as the average \Nc{} recovered from the "25\% Neutral + 75\% Cation" and "Cation" tracks of the charge-size grid \citep[][their Eq. 3]{Maragkoudakis2020} in which the template regions reside. Specifically, the \ANc{} values for the five template regions are \HII{}: \ANc{} = 67, APDR: \ANc{} = 64, DF1: \ANc{} = 62, DF2: \ANc{} = 55, DF3: \ANc{} = 50. The \ANc{} recovered from pyPAHdb modelling, in absence of the 3.3 \micron{} PAH band, shows similar scaling with the $I_{3.3}/I_{11.2}$, i.e. increasing with decreasing $I_{3.3}/I_{11.2}$, as in the charge--size grid (apart from the \HII{} region; see previous paragraph). The \ANc{} values returned from pyPAHdb (Table \ref{tab:templates_breakdown} and Figure  \ref{fig:33_112_vs_Nc_templates}), and those from observational PAH ratio proxies ($I_{3.3}/I_{11.2}$) in conjunction with PAHdb models \citep[e.g.,][]{Maragkoudakis2020, Maragkoudakis2023a, Maragkoudakis2023b} are expected to show differences, given the additional information on the small end of the PAH size distribution obtained from the 3.3 \micron{} band intensity in the latter case and fundamentally using different sets of PAH spectra for both modelling approaches. In addition, extinction may affect the measured $I_{11.2}/I_{3.3}$ intensity ratio, with extinction-corrected values potentially shifting the deduced \ANc{} to lower values in the charge-size diagram. Nevertheless, relative comparison of the \ANc{} recovered by pyPAHdb between different regions, provides a consistent description of the size variation of PAHs among these regions (Figures \ref{fig:33_112_vs_Nc_templates} and \ref{fig:charge-size-grid}).

\section{Discussion} \label{sec:discussion}

The NASA Ames pyPAHdb is a convenient and efficient spectral modelling tool focused on the analysis of JWST spectral mosaics. It provides the means to model the PAH emission spectrum of astronomical sources using emission spectra of individual PAHs. pyPAHdb delivers maps of the different PAH charge states and size components from the PAH molecules contributing to the fit, along with maps of the PAH ionisation parameter $\gamma$ and \ANc{}. 

Utilising pyPAHdb, we have modelled $5-15$ \micron{} PAH emission spectra across the Orion Bar, obtained from the decomposition of the PDRs4All MIRI-MRS mosaic observations, revealing the fractional contribution of different PAH charge states and different sized PAHs to the total PAH emission.

With a larger influx of UV photons at the edge of the PDR, PAH cations are expected to be at a higher concentration in the atomic PDR zone following the IF, gradually decreasing towards the molecular cloud zone as UV photons become attenuated, and, consequently, the concentration of neutral PAHs increases. pyPAHdb modelling supports this picture, where the peak of the fractional contribution of PAH cations is found in the APDR zone and neutral PAHs have minimum contribution. Accordingly, high $f_{neut}$ values are observed in the face-on emission of the \HII{} region and background molecular cloud regions past DF2, where UV photons are sufficiently attenuated to decrease the rate of PAH ionisation. Further corroboration is found from the peak of PAH anion emission, which is observed deep within the DF2 and DF3 zones. The relatively higher fractions of PAH cations (and lower fractions of PAH anions) in BDF3 than toward DF3 further support the hypothesis that a significant component of PAH emission toward BDF3 may arise from layers not directly associated with the Bar itself. 

In regions of high radiation field intensities, photodissociation and photodestruction of small PAHs is expected to actively take place \citep[e.g.][]{Joblin1996, Berne2015, Peeters2024, Chown2024}. pyPAHdb modelling results affirm this, with the peak of the fractional contribution of \Nc{} $\leq 50$ PAHs found between the DF2 and DF3 zones, and their lowest contribution in the IF and \HII{} region. Overall, small and medium sized PAHs make up $\sim 70\%$ of the PAH emission across the mosaic, while $\sim 30\%$ is due to large PAHs. In general, the PAH size distribution will be shaped by the interplay between processes such as destruction, fragmentation, and bottom-up formation of PAH molecules. The magnitude of each process will vary depending on the chemical and physical conditions in the local environment, i.e. PAH destruction will be more prevalent in the \HII{} region and the IF, followed by photodissociation/fragmentation \citep[or top-down formation e.g.][]{Cesarsky2000, Berne2007, Pilleri2012} where larger PAHs, or small carbonaceous grains, are fragmented to produce medium-sized PAHs, which in turn produce smaller PAHs. This succession can be observed at the border of the IF in the PAH size breakdown maps (Figure \ref{fig:size_breakdown_maps}), where: $f_{lar}$ peaks before the IF, $f_{med}$ is high at the PDR right after the IF, and $f_{sm}$ gradually increases within the APDR zone compared to the \HII{} zone and IF. The transient presence of trace amounts of C$_2$H in the atomic PDR \citep[][]{Goicoechea2025} provides further support for photodestruction of PAHs in this region, as laboratory experiments show that photodestruction of PAHs produces C$_2$H$_2$ \citep[][]{Jochims1994, Ekern1998, Zhen2015} and photodissociation of C$_2$H$_2$ produces C$_2$H \citep[][]{Cheng2011, Heays2017}. Within the APDR and up to the DF2 the size distribution is balanced with an even contribution ($30\% - 35\%$) from different sized molecules. Beyond DF2, where photodissociation or fragmentation are less likely to occur due to the drop in the UV intensity, a bottom-up PAH formation may take place, where intermediate and large PAHs are formed, at different rates, from the pool of \Nc{} $\leq 50$ PAHs and by the enhanced abundances of small hydrocarbon radicals \citep[e.g.,][]{Goicoechea2025}. This can be seen in the PAH size breakdown maps (Figure \ref{fig:size_breakdown_maps}) between DF2 and DF3, and especially at DF3.

The derived $\gamma$ across the mosaic ranges between $\sim2 - 9 \times 10^4$ K$^{1/2}$cm$^{-3}$, which fall between the values reported by \cite{Berne2022a} ($0.7 - 2.0 \times 10^4$) K$^{1/2}$cm$^{-3}$ based on model results from \cite{Joblin2018}, and $10.4 \times 10^4$ K$^{1/2}$cm$^{-3}$ derived from the NIRSpec observations \citep{Peeters2024}. At the IF $\gamma$ is on average $4.4 \times 10^4$ K$^{1/2}$cm$^{-3}$, which is in agreement with \cite{Salgado2016} ($\gamma=4\times10^4$ K$^{1/2}$cm$^{-3}$), who assumed constant temperature ($T_{gas}$ = 500 K) and electron density ($n_e$ = 15 cm$^{-3}$), with $G_0$ estimated from the incident radiation field attenuated in the Bar. Similarly at the position of the H$_2$ peak, roughly at DF3, \cite{Salgado2016} reported $\gamma=1.5\times10^4$ K$^{1/2}$cm$^{-3}$ and pyPAHdb analysis derived $\gamma=3.5\times10^4$ K$^{1/2}$cm$^{-3}$. The derived $\gamma$ in this work is dependent on the adopted multiplication factor for the density ratio of cation to neutral PAHs (Eq. \ref{eq:npah_ioniz_param}), which in this case is taken for circumcoronene (\Nc{} = 54). Although the average PAH size across the mosaic is somewhat similar, \ANc{} = 67, applying a single factor for a distribution of PAH sizes may be less appropriate, explaining perhaps some of the variation in $\gamma$ with the previous works. 
 
Currently, pyPAHdb can very successfully model the $5-15$ \micron{} JWST PAH spectrum. Inclusion of the $3.3-3.4$ \micron{} bands for simultaneous modelling of the $3-15$ \micron{} PAH spectrum would allow sampling the carriers of the low end of the PAH size distribution in more detail. Modelling of these shorter wavelength bands at $3.3-3.4$ \micron{} requires implementation of anharmonic vibrational spectroscopy of PAHs \citep{Esposito2024}.\footnote{Anharmonic spectra of PAHs are to be added to PAHdb in a forthcoming release.} The term ``anharmonicity'' as used here refers to the inclusion of higher-order terms (e.g., cubic and quartic) in the vibrational calculation that lead to more accurate vibrational frequencies as well as the treatment of resonances and mode coupling that redistributes band intensities and introduces new bands \citep[e.g. at 5.25 and 5.7 \micron{}][]{Boersma2009}.
Indeed, it has been shown that the positions and profiles of the astronomical 3.3 and 3.4 \micron{} bands can only be reproduced with treatment of anharmonicity in the computations \citep[][]{Mackie2016, Maltseva2018}. Furthermore, the weak, but many, overtone and combination bands produce a so-called PAH-continuum that spans $3-5$ \micron{} and anharmonic results also suggest that the separation of the $3.3-3.4$ \micron{} aromatic/aliphatic PAH features is less straightforward than initially thought \citep[][]{Allamandola2021, Boersma2023, Esposito2024b}. Together, this severely complicates analysing the astronomical $3-6$ \micron{} PAH spectrum. 
Nonetheless, at first order, sampling of the lower end of the PAH size distribution via the 3.3 \micron{} PAH feature would provide additional constraints in the pyPAHdb-derived PAH size determinations.

The $I_{6.2}/I_{11.2}$, $I_{7.7}/I_{11.2}$, and $I_{8.6}/I_{11.2}$ PAH intensity ratios, which are empirical tracers of PAH ionisation, scale extremely well with $\gamma$ (Figure \ref{fig:gamma-pah_ratios}). This correlation is effectively linear for PAH emission in the \HII{} zone (which emanates from the background face-on PDR), the APDR zone, across DF1 and up to DF2. Two separate linear branches are seen in the $\gamma-I_{6.2,7.7}/I_{11.2}$ relations, which include: i) regions in the \HII{} zone and the APDR template (bottom branch); ii) regions in the APDR zone (excluding the APDR template) and up to the DF2, as well as regions beyond DF3 (top branch). In the $\gamma-I_{8.6}/I_{11.2}$ case the two branches are absent (or merged). Such variation in the $\gamma-I_{X}/I_{11.2}$ relations, i.e. the presence (or absence) of branches, is the result of variation in the PAH intensity ratios in the different zones, which could reflect variations in the carriers of the 6.2, 7.7, and 8.6 \micron{} emission, but also variations due to extinction \citep[][]{Peeters2024}. In the molecular cloud zone, between DF2 and DF3, $\gamma$ and $I_{X}/I_{11.2}$ do not correlate. This marks a distinction between PAH ionisation as traced empirically and via spectral modelling, in regions of dominant PDR emission (\HII{}, APDR, and potentially part of the BDF3 zones) and regions within the molecular cloud. Within the DF2-DF3 zones the PAH intensity ratios have the largest variation (Figure \ref{fig:pah_ratios_maps}) when compared to the other zones, whereas $\gamma$ estimated from pyPAHdb modelling has a confined range at lower values (between $3 - 4.5\times10^4$) in comparison to zones closer to the IF. This is a clear case where empirical tracers involving PAH intensity ratios of two bands cannot efficiently describe PAH ionisation, whereas modelling of the entire $5-15$ \micron{} PAH spectrum provides a detailed characterization of the net contribution of neutral and cationic PAHs within the different PAH bands. 

\section{Conclusions} \label{sec:Summary}

Using the NASA Ames pyPAHdb spectral modelling tool we have modelled $5-15$ \micron{} PAH emission spectra across the PDRs4All Orion Bar MIRI-MRS mosaic, including the five key physical zones, i.e., the \HII{} region, the atomic PDR (APDR), and the three bright H$\,\textsc{i}$/H$_{2}$ dissociation fronts (DF1, DF2, and DF3) corresponding to three molecular hydrogen (H$_{2}$) filaments. The main conclusions from our work are summarised below.

(i) Cationic PAH emission peaks in the APDR region (60\% - 70\%), where neutral PAHs have minimal contribution (up to $\sim 25\%$). Emission from neutral PAHs peak in the line-of-sight towards the \HII{} region (40\% - 50\%), originating in a background face-on PDR, and from the underlying molecular cloud regions past DF2 ($\sim 40\%$). PAH anions are observed (up to $\sim 20\%$) deep within the DF2 and DF3 zones. The PAH ionisation parameter $\gamma$ ranges between $\sim 2-9\times10^4$ across the mosaic. 

(ii) Small (\Nc{} $\leq 50$) and medium ($50 <$ \Nc{} $\leq 70$) sized PAHs make up $\sim 70\%$ of the PAH emission across the mosaic, with the peak of the small PAH emission ($45\%-50\%$) found between DF2 and DF3. The average PAH size \Nc{} across the Orion Bar ranges between $\sim 60-74$. 

(iii) The fractional PAH charge and size breakdown from pyPAHdb modelling reveals regions of likely top-down PAH formation at the ionisation front, and bottom-up PAH formation within the molecular cloud region.

(iv) The derived average PAH size \Nc{} in the different physical zones is consistent with a scenario where PAHs are being more extensively subjected to ultraviolet processing closer to the ionisation front, and less within the molecular cloud.

(v) Empirical tracers for PAH ionisation scale well with $\gamma$ in regions dominated by PDR emission, but their correlation weakens within the molecular cloud zone. Therefore, modelling of the $5-15$ \micron{} PAH spectrum with pyPAHdb achieves comprehensive characterization of the net contribution of neutral and cationic PAHs across different environments, whereas empirical PAH intensity ratio tracers can be highly variable and less reliable outside regions dominated by PDR emission.

\begin{acknowledgements}
We would like to thank the referee for providing constructive comments and suggestions that have improved the clarity of this paper. A.M., L.J.A., V.J.E., and J.D.B.'s research was supported by an appointment at NASA Ames Research Center, administered by the Bay Area Environmental Research Institute (80NSSC23M0028). C.B. is grateful for an appointment at NASA Ames Research Center through the San Jos\'e State University Research Foundation (80NSSC22M0107). A.M., C.B., L.J.A., P.T., V.J.E., J.D.B., and A.R. gratefully acknowledge support from the ``NASA Ames Laboratory Astrophysics Directed Work Package (LADWP) Round 2 ISFM'' (22-A22ISFM-0009). E.P. acknowledges support from the University of Western Ontario, the Canadian Space Agency (CSA, 22JWGO1-16), and the Natural Sciences and Engineering Research Council of Canada. J.R.G. thanks the Spanish MCINN for funding support under grant PID2023-146667NB-I00. A.F. acknowledges funding from the European Research Council (ERC) under the European Union’s Horizon Europe research and innovation programme ERC-AdG-2022 (GA No. 101096293)
A.F. also thanks project PID2022-137980NB-I00 funded by the Spanish Ministry of Science and Innovation/State Agency of Research MCIN/AEI/10.13039/501100011033 and by “ERDF A way of making Europe”. M.B. acknowledges DST, India for the 'DST INSPIRE Faculty' fellowship and grant. M.B. also acknowledges IUCAA, Pune for visiting associateship.

\end{acknowledgements}
 


\bibliographystyle{aa}
\bibliography{bibliography}{} 

\begin{appendix}

\section{pyPAHdb modelling error} \label{sec:pypahdb_unc}

\begin{figure}
    \centering
    \includegraphics[scale=0.65]{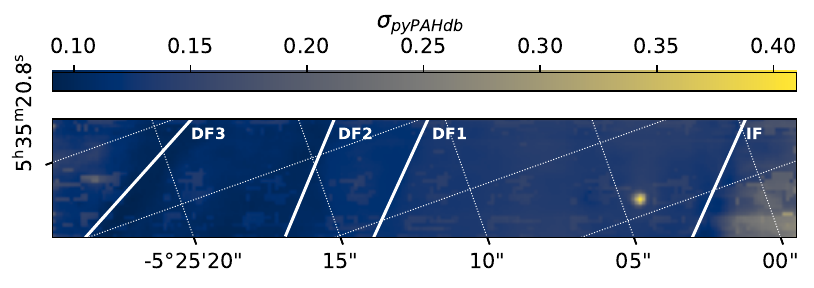}
    \caption{The pyPAHdb modelling error ($\sigma_{pyPAHdb}$) map.}
    \label{fig:pyPAHdb_errormap}
\end{figure}

The pyPAHdb modelling error ($\sigma_{pyPAHdb}$) is quantified as the ratio between the integrals of the absolute residuals over that of the absolute input spectrum \citep[e.g.][]{Bauschlicher2018, Maragkoudakis2022, Maragkoudakis2025}. The map of $\sigma_{pyPAHdb}$ is presented in Figure \ref{fig:pyPAHdb_errormap}. Regions of relatively high $\sigma_{pyPAHdb}$ ($> 0.30$) are seen in the 203-504 proplyd and part of the \HII{} region, whereas the mean $\sigma_{pyPAHdb}$ is 0.13 $\pm$ 0.03, similarly to the average modelling error ($\overline{\sigma_{pyPAHdb}}$ = 0.16) obtained for nearby galaxies \citep[][]{Maragkoudakis2025}.

\section{pyPAHdb modelling of the five key physical zones} \label{sec:template_plots}

\begin{figure}
    \centering
    \includegraphics[scale=0.52]{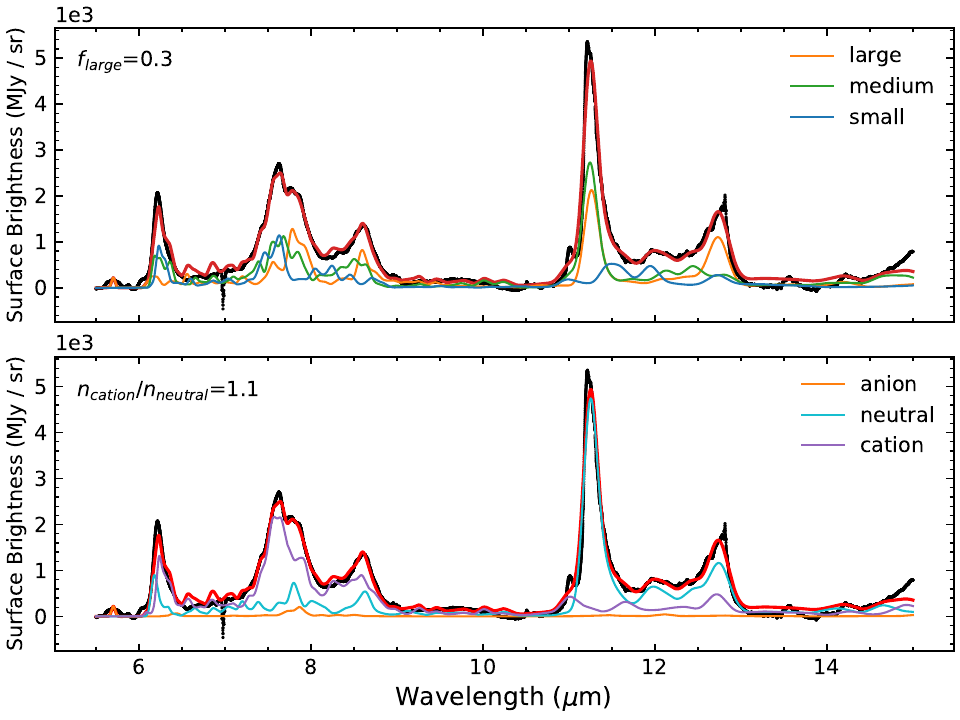}\\
    \includegraphics[scale=0.52]{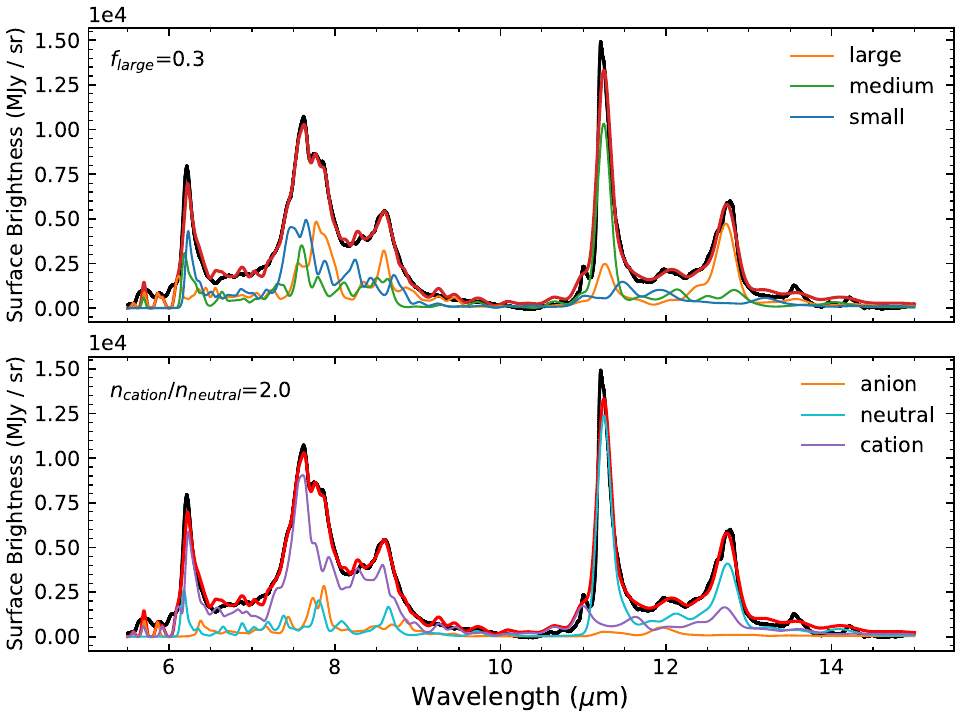} \\
    \includegraphics[scale=0.52]{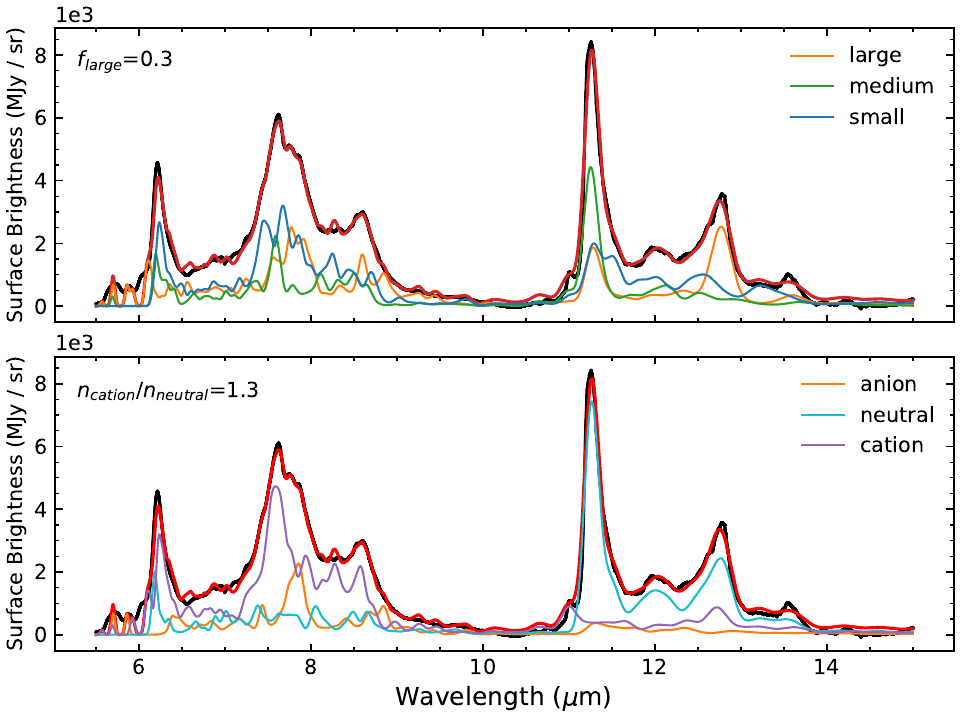} \\
    \caption{The pyPAHdb PAH size and charge breakdown of the spectral templates of the \HII{} (top), DF1 (middle), and DF2 (bottom) regions. Observations are shown in black, total fit in red, and the separate breakdown components are given in the legend of each panel. The fractions of the large PAH contributing to the fit and the cation-to-neutral PAHs ratio are provided as labels.}
    \label{fig:appendix_pyPAHdb_templates}
\end{figure}

Modelling to the weighted averaged (template) spectrum of each of the five key physical zones was performed with pyPAHdb (Section \ref{sec:templates}). The plots of the PAH charge and size breakdown for the APDR and DF3 regions are presented in Figure \ref{fig:pyPAHdb_templates} and for the HII, DF1, and DF3 regions in Figure \ref{fig:appendix_pyPAHdb_templates}.

\end{appendix}

\end{document}

%% file: affiliations.tex
\newcommand{\ames}{1}
\newcommand{\uwo}{2}
\newcommand{\wspace}{3}
\newcommand{\seti}{4}
\newcommand{\umich}{5}
\newcommand{\toulouse}{6}
\newcommand{\uoh}{7}
\newcommand{\osu}{8}
\newcommand{\paris}{9}
\newcommand{\csic}{10}
\newcommand{\ipac}{11}
\newcommand{\tokyoastro}{12}
\newcommand{\maryland}{13}
\newcommand{\stsci}{14}
\newcommand{\sunyatsen}{15}

\newcommand{\amesname}{NASA Ames Research Center, MS 245-6, Moffett Field, CA 94035-1000, USA}
\newcommand{\uwoname}{Department of Physics \& Astronomy, The University of Western Ontario, London ON N6A 3K7, Canada}
\newcommand{\wspacename}{Institute for Earth and Space Exploration, The University of Western Ontario, London ON N6A 3K7, Canada}
\newcommand{\osuname}{Department of Astronomy, The Ohio State University, 140 West 18th Avenue, Columbus, OH 43210, USA }

\newcommand{\setiname}{Carl Sagan Center, SETI Institute, 339 Bernardo Avenue, Suite 200, Mountain View, CA 94043, USA }
\newcommand{\parisname}{Institut d'Astrophysique Spatiale, Université Paris-Saclay, CNRS,  Bâtiment 121, 91405 Orsay Cedex, France}
\newcommand{\toulousename}{Institut de Recherche en Astrophysique et Planétologie, Université Toulouse III - Paul Sabatier, CNRS, CNES, 9 Av. du colonel Roche, 31028 Toulouse Cedex 04, France}
\newcommand{\stsciname}{Space Telescope Science Institute, 3700 San Martin Drive, Baltimore, MD 21218, USA}
\newcommand{\umichname}{Department of Astronomy, University of Michigan, 1085 South University Avenue, Ann Arbor, MI 48109, USA}
\newcommand{\uohname}{School of Physics, University of Hyderabad, Hyderabad, Telangana 500046, India}
\newcommand{\csicname}{Instituto de Física Fundamental (CSIC), Calle Serrano 121-123, 28006 Madrid, Spain}
\newcommand{\ipacname}{IPAC, California Institute of Technology, Pasadena, CA, USA}
\newcommand{\marylandname}{Astronomy Department, University of Maryland, College Park, MD 20742, USA}
\newcommand{\sunyatsenname}{School of Physics and Astronomy, Sun Yat-sen University, 2 Da Xue Road, Tangjia, Zhuhai 519000,  Guangdong Province, China}
\newcommand{\tokyoastroname}{Department of Astronomy, Graduate School of Science, The University of Tokyo, 7-3-1 Bunkyo-ku, Tokyo 113-0033, Japan}

%% file: bibliography.bib
@ARTICLE{Allamandola1985,
   author = {{Allamandola}, L.~J. and {Tielens}, A.~G.~G.~M. and {Barker}, J.~R.
	},
    title = "{Polycyclic aromatic hydrocarbons and the unidentified infrared emission bands - Auto exhaust along the Milky Way}",
  journal = {\apjl},
     year = 1985,
    month = mar,
   volume = 290,
    pages = {L25-L28},
      doi = {10.1086/184435},
   adsurl = {http://adsabs.harvard.edu/abs/1985ApJ...290L..25A}
}

@ARTICLE{Allamandola1989,
   author = {{Allamandola}, L.~J. and {Tielens}, A.~G.~G.~M. and {Barker}, J.~R.
	},
    title = "{Interstellar polycyclic aromatic hydrocarbons - The infrared emission bands, the excitation/emission mechanism, and the astrophysical implications}",
  journal = {\apjs},
     year = 1989,
    month = dec,
   volume = 71,
    pages = {733-775},
      doi = {10.1086/191396},
   adsurl = {http://adsabs.harvard.edu/abs/1989ApJS...71..733A}
}

@ARTICLE{Allamandola1999,
       author = {{Allamandola}, L.~J. and {Hudgins}, D.~M. and {Sandford}, S.~A.},
        title = "{Modeling the Unidentified Infrared Emission with Combinations of Polycyclic Aromatic Hydrocarbons}",
      journal = {\apj},
         year = "1999",
        month = "Feb",
       volume = {511},
       number = {2},
        pages = {L115-L119},
          doi = {10.1086/311843},
       adsurl = {https://ui.adsabs.harvard.edu/abs/1999ApJ...511L.115A}
}

@ARTICLE{Andrews2015,
       author = {{Andrews}, H. and {Boersma}, C. and {Werner}, M.~W. and {Livingston}, J. and {Allamandola}, L.~J. and {Tielens}, A.~G.~G.~M.},
        title = "{PAH Emission at the Bright Locations of PDRs: the grandPAH Hypothesis}",
      journal = {\apj},
         year = 2015,
        month = jul,
       volume = {807},
       number = {1},
          eid = {99},
        pages = {99},
          doi = {10.1088/0004-637X/807/1/99},
       adsurl = {https://ui.adsabs.harvard.edu/abs/2015ApJ...807...99A}
}

@ARTICLE{BakesTielens1994,
   author = {{Bakes}, E.~L.~O. and {Tielens}, A.~G.~G.~M.},
    title = "{The photoelectric heating mechanism for very small graphitic grains and polycyclic aromatic hydrocarbons}",
  journal = {\apj},
     year = 1994,
    month = jun,
   volume = 427,
    pages = {822-838},
      doi = {10.1086/174188},
   adsurl = {http://adsabs.harvard.edu/abs/1994ApJ...427..822B}
}

@ARTICLE{Bauschlicher2010,
	author = {{Bauschlicher}, Jr., C.~W. and {Boersma}, C. and {Ricca}, A. and 
	{Mattioda}, A.~L. and {Cami}, J. and {Peeters}, E. and {S{\'a}nchez de Armas}, F. and 
	{Puerta Saborido}, G. and {Hudgins}, D.~M. and {Allamandola}, L.~J.
	},
	title = "{The NASA Ames Polycyclic Aromatic Hydrocarbon Infrared Spectroscopic Database: The Computed Spectra}",
	journal = {\apjs},
	year = 2010,
	month = aug,
	volume = 189,
	eid = {341},
	pages = {341-351},
	doi = {10.1088/0067-0049/189/2/341},
	adsurl = {http://adsabs.harvard.edu/abs/2010ApJS..189..341B}
}

@ARTICLE{Bauschlicher2018,
       author = {{Bauschlicher}, Charles W., Jr. and {Ricca}, A. and {Boersma}, C. and
         {Allamandola}, L.~J.},
        title = "{The NASA Ames PAH IR Spectroscopic Database: Computational Version 3.00 with Updated Content and the Introduction of Multiple Scaling Factors}",
      journal = {\apjs},
         year = "2018",
        month = "Feb",
       volume = {234},
       number = {2},
          eid = {32},
        pages = {32},
          doi = {10.3847/1538-4365/aaa019},
       adsurl = {https://ui.adsabs.harvard.edu/abs/2018ApJS..234...32B}
}

@ARTICLE{Beintema1996,
       author = {{Beintema}, D.~A. and {van den Ancker}, M.~E. and {Molster}, F.~J. and {Waters}, L.~B.~F.~M. and {Tielens}, A.~G.~G.~M. and {Waelkens}, C. and {de Jong}, T. and {de Graauw}, T. and {Justtanont}, K. and {Yamamura}, I. and {Heras}, A. and {Lahuis}, F. and {Salama}, A.},
        title = "{The rich spectrum of circumstellar PAHs.}",
      journal = {\aap},
         year = 1996,
        month = nov,
       volume = {315},
        pages = {L369-L372},
       adsurl = {https://ui.adsabs.harvard.edu/abs/1996A&A...315L.369B}
}

@ARTICLE{Berne2007,
       author = {{Bern{\'e}}, O. and {Joblin}, C. and {Deville}, Y. and {Smith}, J.~D. and {Rapacioli}, M. and {Bernard}, J.~P. and {Thomas}, J. and {Reach}, W. and {Abergel}, A.},
        title = "{Analysis of the emission of very small dust particles from Spitzer spectro-imagery data using blind signal separation methods}",
      journal = {\aap},
         year = 2007,
        month = jul,
       volume = {469},
       number = {2},
        pages = {575-586},
          doi = {10.1051/0004-6361:20066282},
archivePrefix = {arXiv},
       eprint = {astro-ph/0703072},
 primaryClass = {astro-ph},
       adsurl = {https://ui.adsabs.harvard.edu/abs/2007A&A...469..575B}
}

@ARTICLE{Berne2022a,
       author = {{Bern{\'e}}, O. and {Foschino}, S. and {Jalabert}, F. and {Joblin}, C.},
        title = "{Contribution of polycyclic aromatic hydrocarbon ionization to neutral gas heating in galaxies: model versus observations}",
      journal = {\aap},
         year = 2022,
        month = nov,
       volume = {667},
          eid = {A159},
        pages = {A159},
          doi = {10.1051/0004-6361/202243171},
archivePrefix = {arXiv},
       eprint = {2208.08762},
 primaryClass = {astro-ph.GA},
       adsurl = {https://ui.adsabs.harvard.edu/abs/2022A&A...667A.159B}
}

@ARTICLE{Berne2022b,
       author = {{Bern{\'e}}, Olivier and {Habart}, {\'E}milie and {Peeters}, Els and {Abergel}, Alain and {Bergin}, Edwin A. and {Bernard-Salas}, Jeronimo and {Bron}, Emeric and {Cami}, Jan and {Dartois}, Emmanuel and {Fuente}, Asunci{\'o}n and {Goicoechea}, Javier R. and {Gordon}, Karl D. and {Okada}, Yoko and {Onaka}, Takashi and {Robberto}, Massimo and {R{\"o}llig}, Markus and {Tielens}, Alexander G.~G.~M. and {Vicente}, S{\'\i}lvia and {Wolfire}, Mark G. and {Alarc{\'o}n}, Felipe and {Boersma}, C. and {Canin}, Am{\'e}lie and {Chown}, Ryan and {Dicken}, Daniel and {Languignon}, David and {Le Gal}, Romane and {Pound}, Marc W. and {Trahin}, Boris and {Simmer}, Thomas and {Sidhu}, Ameek and {Van De Putte}, Dries and {Cuadrado}, Sara and {Guilloteau}, Claire and {Maragkoudakis}, Alexandros and {Schefter}, Bethany R. and {Schirmer}, Thi{\'e}baut and {Cazaux}, St{\'e}phanie and {Aleman}, Isabel and {Allamandola}, Louis and {Auchettl}, Rebecca and {Baratta}, Giuseppe Antonio and {Bejaoui}, Salma and {Bera}, Partha P. and {Bilalbegovi{\'c}}, Goranka and {Black}, John H. and {Boulanger}, Francois and {Bouwman}, Jordy and {Brandl}, Bernhard and {Brechignac}, Philippe and {Br{\"u}nken}, Sandra and {Burkhardt}, Andrew and {Candian}, Alessandra and {Cernicharo}, Jose and {Chabot}, Marin and {Chakraborty}, Shubhadip and {Champion}, Jason and {Colgan}, Sean W.~J. and {Cooke}, Ilsa R. and {Coutens}, Audrey and {Cox}, Nick L.~J. and {Demyk}, Karine and {Donovan Meyer}, Jennifer and {Engrand}, C{\'e}cile and {Foschino}, Sacha and {Garc{\'\i}a-Lario}, Pedro and {Gavilan}, Lisseth and {Gerin}, Maryvonne and {Godard}, Marie and {Gottlieb}, Carl A. and {Guillard}, Pierre and {Gusdorf}, Antoine and {Hartigan}, Patrick and {He}, Jinhua and {Herbst}, Eric and {Hornekaer}, Liv and {J{\"a}ger}, Cornelia and {Janot-Pacheco}, Eduardo and {Joblin}, Christine and {Kaufman}, Michael and {Kemper}, Francisca and {Kendrew}, Sarah and {Kirsanova}, Maria S. and {Klaassen}, Pamela and {Knight}, Collin and {Kwok}, Sun and {Labiano}, {\'A}lvaro and {Lai}, Thomas S. -Y. and {Lee}, Timothy J. and {Lefloch}, Bertrand and {Le Petit}, Franck and {Li}, Aigen and {Linz}, Hendrik and {Mackie}, Cameron J. and {Madden}, Suzanne C. and {Mascetti}, Jo{\"e}lle and {McGuire}, Brett A. and {Merino}, Pablo and {Micelotta}, Elisabetta R. and {Misselt}, Karl and {Morse}, Jon A. and {Mulas}, Giacomo and {Neelamkodan}, Naslim and {Ohsawa}, Ryou and {Omont}, Alain and {Paladini}, Roberta and {Palumbo}, Maria Elisabetta and {Pathak}, Amit and {Pendleton}, Yvonne J. and {Petrignani}, Annemieke and {Pino}, Thomas and {Puga}, Elena and {Rangwala}, Naseem and {Rapacioli}, Mathias and {Ricca}, Alessandra and {Roman-Duval}, Julia and {Roser}, Joseph and {Roueff}, Evelyne and {Rouill{\'e}}, Ga{\"e}l and {Salama}, Farid and {Sales}, Dinalva A. and {Sandstrom}, Karin and {Sarre}, Peter and {Sciamma-O'Brien}, Ella and {Sellgren}, Kris and {Shannon}, Matthew J. and {Shenoy}, Sachindev S. and {Teyssier}, David and {Thomas}, Richard D. and {Togi}, Aditya and {Verstraete}, Laurent and {Witt}, Adolf N. and {Wootten}, Alwyn and {Ysard}, Nathalie and {Zettergren}, Henning and {Zhang}, Yong and {Zhang}, Ziwei E. and {Zhen}, Junfeng},
        title = "{PDRs4All: A JWST Early Release Science Program on Radiative Feedback from Massive Stars}",
      journal = {\pasp},
         year = 2022,
        month = may,
       volume = {134},
       number = {1035},
          eid = {054301},
        pages = {054301},
          doi = {10.1088/1538-3873/ac604c},
archivePrefix = {arXiv},
       eprint = {2201.05112},
 primaryClass = {astro-ph.GA},
       adsurl = {https://ui.adsabs.harvard.edu/abs/2022PASP..134e4301B}
}

@ARTICLE{Berne2023,
       author = {{Bern{\'e}}, Olivier and {Martin-Drumel}, Marie-Aline and {Schroetter}, Ilane and {Goicoechea}, Javier R. and {Jacovella}, Ugo and {Gans}, B{\'e}renger and {Dartois}, Emmanuel and {Coudert}, Laurent H. and {Bergin}, Edwin and {Alarcon}, Felipe and {Cami}, Jan and {Roueff}, Evelyne and {Black}, John H. and {Asvany}, Oskar and {Habart}, Emilie and {Peeters}, Els and {Canin}, Amelie and {Trahin}, Boris and {Joblin}, Christine and {Schlemmer}, Stephan and {Thorwirth}, Sven and {Cernicharo}, Jose and {Gerin}, Maryvonne and {Tielens}, Alexander and {Zannese}, Marion and {Abergel}, Alain and {Bernard-Salas}, Jeronimo and {Boersma}, Christiaan and {Bron}, Emeric and {Chown}, Ryan and {Cuadrado}, Sara and {Dicken}, Daniel and {Elyajouri}, Meriem and {Fuente}, Asunci{\'o}n and {Gordon}, Karl D. and {Issa}, Lina and {Kannavou}, Olga and {Khan}, Baria and {Lacinbala}, Ozan and {Languignon}, David and {Le Gal}, Romane and {Maragkoudakis}, Alexandros and {Meshaka}, Raphael and {Okada}, Yoko and {Onaka}, Takashi and {Pasquini}, Sofia and {Pound}, Marc W. and {Robberto}, Massimo and {R{\"o}llig}, Markus and {Schefter}, Bethany and {Schirmer}, Thi{\'e}baut and {Sidhu}, Ameek and {Tabone}, Benoit and {Van De Putte}, Dries and {Vicente}, S{\'\i}lvia and {Wolfire}, Mark G.},
        title = "{Formation of the methyl cation by photochemistry in a protoplanetary disk}",
      journal = {\nat},
         year = 2023,
        month = sep,
       volume = {621},
       number = {7977},
        pages = {56-59},
          doi = {10.1038/s41586-023-06307-x},
archivePrefix = {arXiv},
       eprint = {2401.03296},
 primaryClass = {astro-ph.GA},
       adsurl = {https://ui.adsabs.harvard.edu/abs/2023Natur.621...56B}
}

@ARTICLE{Berne2024,
       author = {{Bern{\'e}}, Olivier and {Habart}, Emilie and {Peeters}, Els and {Schroetter}, Ilane and {Canin}, Am{\'e}lie and {Sidhu}, Ameek and {Chown}, Ryan and {Bron}, Emeric and {Haworth}, Thomas J. and {Klaassen}, Pamela and {Trahin}, Boris and {Van De Putte}, Dries and {Alarc{\'o}n}, Felipe and {Zannese}, Marion and {Abergel}, Alain and {Bergin}, Edwin A. and {Bernard-Salas}, Jeronimo and {Boersma}, Christiaan and {Cami}, Jan and {Cuadrado}, Sara and {Dartois}, Emmanuel and {Dicken}, Daniel and {Elyajouri}, Meriem and {Fuente}, Asunci{\'o}n and {Goicoechea}, Javier R. and {Gordon}, Karl D. and {Issa}, Lina and {Joblin}, Christine and {Kannavou}, Olga and {Khan}, Baria and {Lacinbala}, Ozan and {Languignon}, David and {Le Gal}, Romane and {Maragkoudakis}, Alexandros and {Meshaka}, Raphael and {Okada}, Yoko and {Onaka}, Takashi and {Pasquini}, Sofia and {Pound}, Marc W. and {Robberto}, Massimo and {R{\"o}llig}, Markus and {Schefter}, Bethany and {Schirmer}, Thi{\'e}baut and {Simmer}, Thomas and {Tabone}, Benoit and {Tielens}, Alexander G.~G.~M. and {Vicente}, S{\'\i}lvia and {Wolfire}, Mark G. and {PDRs4All Team} and {Aleman}, Isabel and {Allamandola}, Louis and {Auchettl}, Rebecca and {Baratta}, Giuseppe Antonio and {Baruteau}, Cl{\'e}ment and {Bejaoui}, Salma and {Bera}, Partha P. and {Black}, John H. and {Boulanger}, Francois and {Bouwman}, Jordy and {Brandl}, Bernhard and {Brechignac}, Philippe and {Br{\"u}nken}, Sandra and {Buragohain}, Mridusmita and {Burkhardt}, Andrew and {Candian}, Alessandra and {Cazaux}, St{\'e}phanie and {Cernicharo}, Jose and {Chabot}, Marin and {Chakraborty}, Shubhadip and {Champion}, Jason and {Colgan}, Sean W.~J. and {Cooke}, Ilsa R. and {Coutens}, Audrey and {Cox}, Nick L.~J. and {Demyk}, Karine and {Meyer}, Jennifer Donovan and {Engrand}, C{\'e}cile and {Foschino}, Sacha and {Garc{\'\i}a-Lario}, Pedro and {Gavilan}, Lisseth and {Gerin}, Maryvonne and {Godard}, Marie and {Gottlieb}, Carl A. and {Guillard}, Pierre and {Gusdorf}, Antoine and {Hartigan}, Patrick and {He}, Jinhua and {Herbst}, Eric and {Hornekaer}, Liv and {J{\"a}ger}, Cornelia and {Janot-Pacheco}, Eduardo and {Kaufman}, Michael and {Kemper}, Francisca and {Kendrew}, Sarah and {Kirsanova}, Maria S. and {Knight}, Collin and {Kwok}, Sun and {Labiano}, {\'A}lvaro and {Lai}, Thomas S. -Y. and {Lee}, Timothy J. and {Lefloch}, Bertrand and {Le Petit}, Franck and {Li}, Aigen and {Linz}, Hendrik and {Mackie}, Cameron J. and {Madden}, Suzanne C. and {Mascetti}, Jo{\"e}lle and {McGuire}, Brett A. and {Merino}, Pablo and {Micelotta}, Elisabetta R. and {Morse}, Jon A. and {Mulas}, Giacomo and {Neelamkodan}, Naslim and {Ohsawa}, Ryou and {Paladini}, Roberta and {Palumbo}, Maria Elisabetta and {Pathak}, Amit and {Pendleton}, Yvonne J. and {Petrignani}, Annemieke and {Pino}, Thomas and {Puga}, Elena and {Rangwala}, Naseem and {Rapacioli}, Mathias and {Ricca}, Alessandra and {Roman-Duval}, Julia and {Roueff}, Evelyne and {Rouill{\'e}}, Ga{\"e}l and {Salama}, Farid and {Sales}, Dinalva A. and {Sandstrom}, Karin and {Sarre}, Peter and {Sciamma-O'Brien}, Ella and {Sellgren}, Kris and {Shannon}, Matthew J. and {Simonnin}, Adrien and {Shenoy}, Sachindev S. and {Teyssier}, David and {Thomas}, Richard D. and {Togi}, Aditya and {Verstraete}, Laurent and {Witt}, Adolf N. and {Wootten}, Alwyn and {Ysard}, Nathalie and {Zettergren}, Henning and {Zhang}, Yong and {Zhang}, Ziwei E. and {Zhen}, Junfeng},
        title = "{A far-ultraviolet{\textendash}driven photoevaporation flow observed in a protoplanetary disk}",
      journal = {Science},
         year = 2024,
        month = mar,
       volume = {383},
       number = {6686},
        pages = {988-992},
          doi = {10.1126/science.adh2861},
archivePrefix = {arXiv},
       eprint = {2403.00160},
 primaryClass = {astro-ph.GA},
       adsurl = {https://ui.adsabs.harvard.edu/abs/2024Sci...383..988B}
}

@ARTICLE{Boersma2009,
   journal = {\apj},
     month = jan,
     pages = {1208-1221},
    author = {{Boersma}, C. and {Mattioda}, A.~L. and {Bauschlicher}, C.~W. and {Peeters}, E. and {Tielens}, A.~G.~G.~M. and {Allamandola}, L.~J.},
       doi = {10.1088/0004-637x/694/1/704},
      year = 2009,
    volume = 690,
     title = "{The 5.25 and 5.7 {$\mu$}m Astronomical Polycyclic Aromatic Hydrocarbon Emission Features}"
}

@ARTICLE{Boersma2013,
       author = {{Boersma}, C. and {Bregman}, J.~D. and {Allamandola}, L.~J.},
        title = "{Properties of Polycyclic Aromatic Hydrocarbons in the Northwest Photon Dominated Region of NGC 7023. I. PAH Size, Charge, Composition, and Structure Distribution}",
      journal = {\apj},
         year = 2013,
        month = jun,
       volume = {769},
       number = {2},
          eid = {117},
        pages = {117},
          doi = {10.1088/0004-637X/769/2/117},
       adsurl = {https://ui.adsabs.harvard.edu/abs/2013ApJ...769..117B}
}

@ARTICLE{Boersma2014a,
	author = {{Boersma}, C. and {Bauschlicher}, Jr., C.~W. and {Ricca}, A. and 
	{Mattioda}, A.~L. and {Cami}, J. and {Peeters}, E. and {S{\'a}nchez de Armas}, F. and 
	{Puerta Saborido}, G. and {Hudgins}, D.~M. and {Allamandola}, L.~J.
	},
	title = "{The NASA Ames PAH IR Spectroscopic Database Version 2.00: Updated Content, Web Site, and On(Off)line Tools}",
	journal = {\apjs},
	year = 2014,
	month = mar,
	volume = 211,
	eid = {8},
	pages = {8},
	doi = {10.1088/0067-0049/211/1/8},
	adsurl = {http://adsabs.harvard.edu/abs/2014ApJS..211....8B}
}

@ARTICLE{Boersma2015,
       author = {{Boersma}, C. and {Bregman}, J. and {Allamandola}, L.~J.},
        title = "{Properties of Polycyclic Aromatic Hydrocarbons in the Northwest Photon Dominated Region of NGC 7023. III. Quantifying the Traditional Proxy for PAH Charge and Assessing its Role}",
      journal = {\apj},
         year = 2015,
        month = jun,
       volume = {806},
       number = {1},
          eid = {121},
        pages = {121},
          doi = {10.1088/0004-637X/806/1/121},
       adsurl = {https://ui.adsabs.harvard.edu/abs/2015ApJ...806..121B}
}

@ARTICLE{Boersma2018,
       author = {{Boersma}, C. and {Bregman}, J. and {Allamandola}, L.~J.},
        title = "{The Charge State of Polycyclic Aromatic Hydrocarbons across a Reflection Nebula, an H II Region, and a Planetary Nebula}",
      journal = {\apj},
         year = 2018,
        month = may,
       volume = {858},
       number = {2},
          eid = {67},
        pages = {67},
          doi = {10.3847/1538-4357/aabcbe},
       adsurl = {https://ui.adsabs.harvard.edu/abs/2018ApJ...858...67B}
}

@ARTICLE{Bregman1989,
       author = {{Bregman}, J.~D. and {Allamandola}, L.~J. and {Tielens}, A.~G.~G.~M. and
         {Geballe}, T.~R. and {Witteborn}, F.~C.},
        title = "{The Infrared Emission Bands. II. A Spatial and Spectral Study of the Orion Bar}",
      journal = {\apj},
         year = "1989",
        month = "Sep",
       volume = {344},
        pages = {791},
          doi = {10.1086/167844},
       adsurl = {https://ui.adsabs.harvard.edu/abs/1989ApJ...344..791B}
}

@ARTICLE{Chown2024,
       author = {{Chown}, Ryan and {Sidhu}, Ameek and {Peeters}, Els and {Tielens}, Alexander G.~G.~M. and {Cami}, Jan and {Bern{\'e}}, Olivier and {Habart}, Emilie and {Alarc{\'o}n}, Felipe and {Canin}, Am{\'e}lie and {Schroetter}, Ilane and {Trahin}, Boris and {Van De Putte}, Dries and {Abergel}, Alain and {Bergin}, Edwin A. and {Bernard-Salas}, Jeronimo and {Boersma}, Christiaan and {Bron}, Emeric and {Cuadrado}, Sara and {Dartois}, Emmanuel and {Dicken}, Daniel and {El-Yajouri}, Meriem and {Fuente}, Asunci{\'o}n and {Goicoechea}, Javier R. and {Gordon}, Karl D. and {Issa}, Lina and {Joblin}, Christine and {Kannavou}, Olga and {Khan}, Baria and {Lacinbala}, Ozan and {Languignon}, David and {Le Gal}, Romane and {Maragkoudakis}, Alexandros and {Meshaka}, Raphael and {Okada}, Yoko and {Onaka}, Takashi and {Pasquini}, Sofia and {Pound}, Marc W. and {Robberto}, Massimo and {R{\"o}llig}, Markus and {Schefter}, Bethany and {Schirmer}, Thi{\'e}baut and {Vicente}, S{\'\i}lvia and {Wolfire}, Mark G. and {Zannese}, Marion and {Aleman}, Isabel and {Allamandola}, Louis and {Auchettl}, Rebecca and {Baratta}, Giuseppe Antonio and {Bejaoui}, Salma and {Bera}, Partha P. and {Black}, John H. and {Boulanger}, Fran{\c{c}}ois and {Bouwman}, Jordy and {Brandl}, Bernhard and {Brechignac}, Philippe and {Br{\"u}nken}, Sandra and {Buragohain}, Mridusmita and {Burkhardt}, Andrew and {Candian}, Alessandra and {Cazaux}, St{\'e}phanie and {Cernicharo}, Jose and {Chabot}, Marin and {Chakraborty}, Shubhadip and {Champion}, Jason and {Colgan}, Sean W.~J. and {Cooke}, Ilsa R. and {Coutens}, Audrey and {Cox}, Nick L.~J. and {Demyk}, Karine and {Meyer}, Jennifer Donovan and {Foschino}, Sacha and {Garc{\'\i}a-Lario}, Pedro and {Gavilan}, Lisseth and {Gerin}, Maryvonne and {Gottlieb}, Carl A. and {Guillard}, Pierre and {Gusdorf}, Antoine and {Hartigan}, Patrick and {He}, Jinhua and {Herbst}, Eric and {Hornekaer}, Liv and {J{\"a}ger}, Cornelia and {Janot-Pacheco}, Eduardo and {Kaufman}, Michael and {Kemper}, Francisca and {Kendrew}, Sarah and {Kirsanova}, Maria S. and {Klaassen}, Pamela and {Kwok}, Sun and {Labiano}, {\'A}lvaro and {Lai}, Thomas S. -Y. and {Lee}, Timothy J. and {Lefloch}, Bertrand and {Le Petit}, Franck and {Li}, Aigen and {Linz}, Hendrik and {Mackie}, Cameron J. and {Madden}, Suzanne C. and {Mascetti}, Jo{\"e}lle and {McGuire}, Brett A. and {Merino}, Pablo and {Micelotta}, Elisabetta R. and {Misselt}, Karl and {Morse}, Jon A. and {Mulas}, Giacomo and {Neelamkodan}, Naslim and {Ohsawa}, Ryou and {Omont}, Alain and {Paladini}, Roberta and {Palumbo}, Maria Elisabetta and {Pathak}, Amit and {Pendleton}, Yvonne J. and {Petrignani}, Annemieke and {Pino}, Thomas and {Puga}, Elena and {Rangwala}, Naseem and {Rapacioli}, Mathias and {Ricca}, Alessandra and {Roman-Duval}, Julia and {Roser}, Joseph and {Roueff}, Evelyne and {Rouill{\'e}}, Ga{\"e}l and {Salama}, Farid and {Sales}, Dinalva A. and {Sandstrom}, Karin and {Sarre}, Peter and {Sciamma-O'Brien}, Ella and {Sellgren}, Kris and {Shenoy}, Sachindev S. and {Teyssier}, David and {Thomas}, Richard D. and {Togi}, Aditya and {Verstraete}, Laurent and {Witt}, Adolf N. and {Wootten}, Alwyn and {Zettergren}, Henning and {Zhang}, Yong and {Zhang}, Ziwei E. and {Zhen}, Junfeng},
        title = "{PDRs4All. IV. An embarrassment of riches: Aromatic infrared bands in the Orion Bar}",
      journal = {\aap},
         year = 2024,
        month = may,
       volume = {685},
          eid = {A75},
        pages = {A75},
          doi = {10.1051/0004-6361/202346662},
archivePrefix = {arXiv},
       eprint = {2308.16733},
 primaryClass = {astro-ph.GA},
       adsurl = {https://ui.adsabs.harvard.edu/abs/2024A&A...685A..75C}
}

@ARTICLE{Cesarsky2000,
   author = {{Cesarsky}, D. and {Jones}, A.~P. and {Lequeux}, J. and {Verstraete}, L.
	},
    title = "{Silicate emission in Orion}",
  journal = {\aap},
   eprint = {astro-ph/0002282},
     year = 2000,
    month = jun,
   volume = 358,
    pages = {708-716},
   adsurl = {http://adsabs.harvard.edu/abs/2000A%26A...358..708C}
}

@ARTICLE{Croiset2016,
       author = {{Croiset}, B.~A. and {Candian}, A. and {Bern{\'e}}, O. and {Tielens},
        A.~G.~G.~M.},
        title = "{Mapping PAH sizes in NGC 7023 with SOFIA}",
      journal = {\aap},
         year = 2016,
        month = May,
       volume = {590},
          eid = {A26},
        pages = {A26},
          doi = {10.1051/0004-6361/201527714},
       adsurl = {https://ui.adsabs.harvard.edu/#abs/2016A&A...590A..26C}
}

@ARTICLE{Draine2007,
       author = {{Draine}, B.~T. and {Li}, Aigen},
        title = "{Infrared Emission from Interstellar Dust. IV. The Silicate-Graphite-PAH Model in the Post-Spitzer Era}",
      journal = {\apj},
         year = 2007,
        month = mar,
       volume = {657},
       number = {2},
        pages = {810-837},
          doi = {10.1086/511055},
archivePrefix = {arXiv},
       eprint = {astro-ph/0608003},
 primaryClass = {astro-ph},
       adsurl = {https://ui.adsabs.harvard.edu/abs/2007ApJ...657..810D}
}

@ARTICLE{Draine2021,
       author = {{Draine}, B.~T. and {Li}, Aigen and {Hensley}, Brandon S. and {Hunt}, L.~K. and {Sandstrom}, K. and {Smith}, J. -D.~T.},
        title = "{Excitation of Polycyclic Aromatic Hydrocarbon Emission: Dependence on Size Distribution, Ionization, and Starlight Spectrum and Intensity}",
      journal = {\apj},
         year = 2021,
        month = aug,
       volume = {917},
       number = {1},
          eid = {3},
        pages = {3},
          doi = {10.3847/1538-4357/abff51},
archivePrefix = {arXiv},
       eprint = {2011.07046},
 primaryClass = {astro-ph.GA},
       adsurl = {https://ui.adsabs.harvard.edu/abs/2021ApJ...917....3D}
}

@ARTICLE{Galliano2008,
   author = {{Galliano}, F. and {Madden}, S.~C. and {Tielens}, A.~G.~G.~M. and 
	{Peeters}, E. and {Jones}, A.~P.},
    title = "{Variations of the Mid-IR Aromatic Features inside and among Galaxies}",
  journal = {\apj},
archivePrefix = "arXiv",
   eprint = {0801.4955},
     year = 2008,
    month = may,
   volume = 679,
    pages = {310-345},
      doi = {10.1086/587051},
   adsurl = {http://adsabs.harvard.edu/abs/2008ApJ...679..310G}
}

@ARTICLE{Habart2024,
       author = {{Habart}, Emilie and {Peeters}, Els and {Bern{\'e}}, Olivier and {Trahin}, Boris and {Canin}, Am{\'e}lie and {Chown}, Ryan and {Sidhu}, Ameek and {Van De Putte}, Dries and {Alarc{\'o}n}, Felipe and {Schroetter}, Ilane and {Dartois}, Emmanuel and {Vicente}, S{\'\i}lvia and {Abergel}, Alain and {Bergin}, Edwin A. and {Bernard-Salas}, Jeronimo and {Boersma}, Christiaan and {Bron}, Emeric and {Cami}, Jan and {Cuadrado}, Sara and {Dicken}, Daniel and {Elyajouri}, Meriem and {Fuente}, Asunci{\'o}n and {Goicoechea}, Javier R. and {Gordon}, Karl D. and {Issa}, Lina and {Joblin}, Christine and {Kannavou}, Olga and {Khan}, Baria and {Lacinbala}, Ozan and {Languignon}, David and {Le Gal}, Romane and {Maragkoudakis}, Alexandros and {Meshaka}, Raphael and {Okada}, Yoko and {Onaka}, Takashi and {Pasquini}, Sofia and {Pound}, Marc W. and {Robberto}, Massimo and {R{\"o}llig}, Markus and {Schefter}, Bethany and {Schirmer}, Thi{\'e}baut and {Tabone}, Benoit and {Tielens}, Alexander G.~G.~M. and {Wolfire}, Mark G. and {Zannese}, Marion and {Ysard}, Nathalie and {Miville-Deschenes}, Marc-Antoine and {Aleman}, Isabel and {Allamandola}, Louis and {Auchettl}, Rebecca and {Baratta}, Giuseppe Antonio and {Bejaoui}, Salma and {Bera}, Partha P. and {Black}, John H. and {Boulanger}, Francois and {Bouwman}, Jordy and {Brandl}, Bernhard and {Brechignac}, Philippe and {Br{\"u}nken}, Sandra and {Buragohain}, Mridusmita and {Burkhardt}, Andrew and {Candian}, Alessandra and {Cazaux}, St{\'e}phanie and {Cernicharo}, Jose and {Chabot}, Marin and {Chakraborty}, Shubhadip and {Champion}, Jason and {Colgan}, Sean W.~J. and {Cooke}, Ilsa R. and {Coutens}, Audrey and {Cox}, Nick L.~J. and {Demyk}, Karine and {Meyer}, Jennifer Donovan and {Foschino}, Sacha and {Garc{\'\i}a-Lario}, Pedro and {Gavilan}, Lisseth and {Gerin}, Maryvonne and {Gottlieb}, Carl A. and {Guillard}, Pierre and {Gusdorf}, Antoine and {Hartigan}, Patrick and {He}, Jinhua and {Herbst}, Eric and {Hornekaer}, Liv and {J{\"a}ger}, Cornelia and {Janot-Pacheco}, Eduardo and {Kaufman}, Michael and {Kemper}, Francisca and {Kendrew}, Sarah and {Kirsanova}, Maria S. and {Klaassen}, Pamela and {Kwok}, Sun and {Labiano}, {\'A}lvaro and {Lai}, Thomas S. -Y. and {Lee}, Timothy J. and {Lefloch}, Bertrand and {Le Petit}, Franck and {Li}, Aigen and {Linz}, Hendrik and {Mackie}, Cameron J. and {Madden}, Suzanne C. and {Mascetti}, Jo{\"e}lle and {McGuire}, Brett A. and {Merino}, Pablo and {Micelotta}, Elisabetta R. and {Misselt}, Karl and {Morse}, Jon A. and {Mulas}, Giacomo and {Neelamkodan}, Naslim and {Ohsawa}, Ryou and {Omont}, Alain and {Paladini}, Roberta and {Palumbo}, Maria Elisabetta and {Pathak}, Amit and {Pendleton}, Yvonne J. and {Petrignani}, Annemieke and {Pino}, Thomas and {Puga}, Elena and {Rangwala}, Naseem and {Rapacioli}, Mathias and {Ricca}, Alessandra and {Roman-Duval}, Julia and {Roser}, Joseph and {Roueff}, Evelyne and {Rouill{\'e}}, Ga{\"e}l and {Salama}, Farid and {Sales}, Dinalva A. and {Sandstrom}, Karin and {Sarre}, Peter and {Sciamma-O'Brien}, Ella and {Sellgren}, Kris and {Shenoy}, Sachindev S. and {Teyssier}, David and {Thomas}, Richard D. and {Togi}, Aditya and {Verstraete}, Laurent and {Witt}, Adolf N. and {Wootten}, Alwyn and {Zettergren}, Henning and {Zhang}, Yong and {Zhang}, Ziwei E. and {Zhen}, Junfeng},
        title = "{PDRs4All. II. JWST's NIR and MIR imaging view of the Orion Nebula}",
      journal = {\aap},
         year = 2024,
        month = may,
       volume = {685},
          eid = {A73},
        pages = {A73},
          doi = {10.1051/0004-6361/202346747},
archivePrefix = {arXiv},
       eprint = {2308.16732},
 primaryClass = {astro-ph.GA},
       adsurl = {https://ui.adsabs.harvard.edu/abs/2024A&A...685A..73H}
}

@ARTICLE{Hony2001,
   author = {{Hony}, S. and {Van Kerckhoven}, C. and {Peeters}, E. and {Tielens}, A.~G.~G.~M. and 
	{Hudgins}, D.~M. and {Allamandola}, L.~J.},
    title = "{The CH out-of-plane bending modes of PAH molecules in astrophysical environments}",
  journal = {\aap},
   eprint = {astro-ph/0103035},
     year = 2001,
    month = may,
   volume = 370,
    pages = {1030-1043},
      doi = {10.1051/0004-6361:20010242},
   adsurl = {http://adsabs.harvard.edu/abs/2001A%26A...370.1030H}
}

@ARTICLE{Leger1984,
   author = {{Leger}, A. and {Puget}, J.~L.},
    title = "{Identification of the 'unidentified' IR emission features of interstellar dust?}",
  journal = {\aap},
     year = 1984,
    month = aug,
   volume = 137,
    pages = {L5-L8},
   adsurl = {http://adsabs.harvard.edu/abs/1984A%26A...137L...5L}
}

@ARTICLE{Maragkoudakis2018a,
	author = {{Maragkoudakis}, A. and {Ivkovich}, N. and {Peeters}, E. and
	{Stock}, D.~J. and {Hemachandra}, D. and {Tielens}, A.~G.~G.~M.},
	title = "{PAHs and star formation in the H II regions of nearby galaxies M83 and M33}",
	journal = {\mnras},
	year = "2018",
	month = "Dec",
	volume = {481},
	pages = {5370-5393},
	doi = {10.1093/mnras/sty2658},
	archivePrefix = {arXiv},
	eprint = {1809.10136},
	primaryClass = {astro-ph.GA},
	adsurl = {https://ui.adsabs.harvard.edu/\#abs/2018MNRAS.481.5370M}
}

@ARTICLE{Maragkoudakis2020,
       author = {{Maragkoudakis}, A. and {Peeters}, E. and {Ricca}, A.},
        title = "{Probing the size and charge of polycyclic aromatic hydrocarbons}",
      journal = {\mnras},
         year = 2020,
        month = may,
       volume = {494},
       number = {1},
        pages = {642-664},
          doi = {10.1093/mnras/staa681},
archivePrefix = {arXiv},
       eprint = {2003.02823},
 primaryClass = {astro-ph.GA},
       adsurl = {https://ui.adsabs.harvard.edu/abs/2020MNRAS.494..642M}
}

@ARTICLE{Maragkoudakis2022,
       author = {{Maragkoudakis}, A. and {Boersma}, C. and {Temi}, P. and {Bregman}, J.~D. and {Allamandola}, L.~J.},
        title = "{Linking Characteristics of the Polycyclic Aromatic Hydrocarbon Population with Galaxy Properties: A Quantitative Approach Using the NASA Ames PAH IR Spectroscopic Database}",
      journal = {\apj},
         year = 2022,
        month = may,
       volume = {931},
       number = {1},
          eid = {38},
        pages = {38},
          doi = {10.3847/1538-4357/ac666f},
archivePrefix = {arXiv},
       eprint = {2204.05292},
 primaryClass = {astro-ph.GA},
       adsurl = {https://ui.adsabs.harvard.edu/abs/2022ApJ...931...38M}
}

@ARTICLE{Maragkoudakis2023a,
       author = {{Maragkoudakis}, A. and {Peeters}, E. and {Ricca}, A.},
        title = "{Spectral variations among different scenarios of PAH processing or formation}",
      journal = {\mnras},
         year = 2023,
        month = apr,
       volume = {520},
       number = {4},
        pages = {5354-5372},
          doi = {10.1093/mnras/stad465},
archivePrefix = {arXiv},
       eprint = {2302.03678},
 primaryClass = {astro-ph.GA},
       adsurl = {https://ui.adsabs.harvard.edu/abs/2023MNRAS.520.5354M}
}

@ARTICLE{Maragkoudakis2023b,
       author = {{Maragkoudakis}, A. and {Peeters}, E. and {Ricca}, A. and {Boersma}, C.},
        title = "{Polycyclic aromatic hydrocarbon size tracers}",
      journal = {\mnras},
         year = 2023,
        month = sep,
       volume = {524},
       number = {3},
        pages = {3429-3436},
          doi = {10.1093/mnras/stad2062},
archivePrefix = {arXiv},
       eprint = {2307.03743},
 primaryClass = {astro-ph.GA},
       adsurl = {https://ui.adsabs.harvard.edu/abs/2023MNRAS.524.3429M}
}

@ARTICLE{Maragkoudakis2025,
       author = {{Maragkoudakis}, A. and {Boersma}, C. and {Temi}, P. and {Bregman}, J.~D. and {Allamandola}, L.~J. and {Esposito}, V.~J. and {Ricca}, A. and {Peeters}, E.},
        title = "{A Sensitivity Analysis of the Modeling of Polycyclic Aromatic Hydrocarbon Emission in Galaxies}",
      journal = {\apj},
         year = 2025,
        month = jan,
       volume = {979},
       number = {1},
          eid = {90},
        pages = {90},
          doi = {10.3847/1538-4357/ad9918},
       adsurl = {https://ui.adsabs.harvard.edu/abs/2025ApJ...979...90M}
}

@ARTICLE{Mattioda2020,
   journal = {\apjs},
    author = {{Mattioda}, A.~L. and {Hudgins}, D.~M. and {Boersma}, C. and {Bauschlicher}, C.~W. and {Ricca}, A. and {Cami}, J. and {Peeters}, E. and {S{\'{a}}nchez de Armas}, F. and {Puerta Saborido}, G. and {Allamandola}, L.~J.},
       doi = {10.3847/1538-4365/abc2c8},
    number = 2,
     month = dec,
     pages = 22,
     title = "{The NASA Ames PAH IR Spectroscopic Database: The Laboratory Spectra}",
    volume = 251,
      year = 2020,
 publisher = {American Astronomical Society}
}

@ARTICLE{Peeters2002a,
       author = {{Peeters}, E. and {Hony}, S. and {Van Kerckhoven}, C. and
         {Tielens}, A.~G.~G.~M. and {Allamandola}, L.~J. and {Hudgins}, D.~M. and
         {Bauschlicher}, C.~W.},
        title = "{The rich 6 to 9 vec mu m spectrum of interstellar PAHs}",
      journal = {\aap},
         year = "2002",
        month = "Aug",
       volume = {390},
        pages = {1089-1113},
          doi = {10.1051/0004-6361:20020773},
archivePrefix = {arXiv},
       eprint = {astro-ph/0205400},
 primaryClass = {astro-ph},
       adsurl = {https://ui.adsabs.harvard.edu/abs/2002A&A...390.1089P}
}

@ARTICLE{Peeters2002b,
       author = {{Peeters}, E. and {Mart{\'\i}n-Hern{\'a}ndez}, N.~L. and {Damour}, F. and {Cox}, P. and {Roelfsema}, P.~R. and {Baluteau}, J. -P. and {Tielens}, A.~G.~G.~M. and {Churchwell}, E. and {Kessler}, M.~F. and {Mathis}, J.~S. and {Morisset}, C. and {Schaerer}, D.},
        title = "{ISO spectroscopy of compact H II regions in the Galaxy. I. The catalogue}",
      journal = {\aap},
         year = 2002,
        month = jan,
       volume = {381},
        pages = {571-605},
          doi = {10.1051/0004-6361:20011516},
       adsurl = {https://ui.adsabs.harvard.edu/abs/2002A&A...381..571P}
}

@ARTICLE{Peeters2004,
   author = {{Peeters}, E. and {Spoon}, H.~W.~W. and {Tielens}, A.~G.~G.~M.
	},
    title = "{Polycyclic Aromatic Hydrocarbons as a Tracer of Star Formation?}",
  journal = {\apj},
   eprint = {astro-ph/0406183},
     year = 2004,
    month = oct,
   volume = 613,
    pages = {986-1003},
      doi = {10.1086/423237},
   adsurl = {http://adsabs.harvard.edu/abs/2004ApJ...613..986P}
}

@ARTICLE{Peeters2017,
	author = {{Peeters}, E. and {Bauschlicher}, Jr., C.~W. and {Allamandola}, L.~J. and 
	{Tielens}, A.~G.~G.~M. and {Ricca}, A. and {Wolfire}, M.~G.},
	title = "{The PAH Emission Characteristics of the Reflection Nebula NGC 2023}",
	journal = {\apj},
	archivePrefix = "arXiv",
	eprint = {1701.06585},
	year = 2017,
	month = feb,
	volume = 836,
	eid = {198},
	pages = {198},
	doi = {10.3847/1538-4357/836/2/198},
	adsurl = {http://adsabs.harvard.edu/abs/2017ApJ...836..198P}
}

@ARTICLE{Peeters2024,
       author = {{Peeters}, Els and {Habart}, Emilie and {Bern{\'e}}, Olivier and {Sidhu}, Ameek and {Chown}, Ryan and {Van De Putte}, Dries and {Trahin}, Boris and {Schroetter}, Ilane and {Canin}, Am{\'e}lie and {Alarc{\'o}n}, Felipe and {Schefter}, Bethany and {Khan}, Baria and {Pasquini}, Sofia and {Tielens}, Alexander G.~G.~M. and {Wolfire}, Mark G. and {Dartois}, Emmanuel and {Goicoechea}, Javier R. and {Maragkoudakis}, Alexandros and {Onaka}, Takashi and {Pound}, Marc W. and {Vicente}, S{\'\i}lvia and {Abergel}, Alain and {Bergin}, Edwin A. and {Bernard-Salas}, Jeronimo and {Boersma}, Christiaan and {Bron}, Emeric and {Cami}, Jan and {Cuadrado}, Sara and {Dicken}, Daniel and {Elyajouri}, Meriem and {Fuente}, Asunci{\'o}n and {Gordon}, Karl D. and {Issa}, Lina and {Joblin}, Christine and {Kannavou}, Olga and {Lacinbala}, Ozan and {Languignon}, David and {Le Gal}, Romane and {Meshaka}, Raphael and {Okada}, Yoko and {Robberto}, Massimo and {R{\"o}llig}, Markus and {Schirmer}, Thi{\'e}baut and {Tabone}, Benoit and {Zannese}, Marion and {Aleman}, Isabel and {Allamandola}, Louis and {Auchettl}, Rebecca and {Baratta}, Giuseppe Antonio and {Bejaoui}, Salma and {Bera}, Partha P. and {Black}, John H. and {Boulanger}, Francois and {Bouwman}, Jordy and {Brandl}, Bernhard and {Brechignac}, Philippe and {Br{\"u}nken}, Sandra and {Buragohain}, Mridusmita and {Burkhardt}, Andrew and {Candian}, Alessandra and {Cazaux}, St{\'e}phanie and {Cernicharo}, Jose and {Chabot}, Marin and {Chakraborty}, Shubhadip and {Champion}, Jason and {Colgan}, Sean W.~J. and {Cooke}, Ilsa R. and {Coutens}, Audrey and {Cox}, Nick L.~J. and {Demyk}, Karine and {Meyer}, Jennifer Donovan and {Foschino}, Sacha and {Garc{\'\i}a-Lario}, Pedro and {Gerin}, Maryvonne and {Gottlieb}, Carl A. and {Guillard}, Pierre and {Gusdorf}, Antoine and {Hartigan}, Patrick and {He}, Jinhua and {Herbst}, Eric and {Hornekaer}, Liv and {J{\"a}ger}, Cornelia and {Janot-Pacheco}, Eduardo and {Kaufman}, Michael and {Kendrew}, Sarah and {Kirsanova}, Maria S. and {Klaassen}, Pamela and {Kwok}, Sun and {Labiano}, {\'A}lvaro and {Lai}, Thomas S. -Y. and {Lee}, Timothy J. and {Lefloch}, Bertrand and {Le Petit}, Franck and {Li}, Aigen and {Linz}, Hendrik and {Mackie}, Cameron J. and {Madden}, Suzanne C. and {Mascetti}, Jo{\"e}lle and {McGuire}, Brett A. and {Merino}, Pablo and {Micelotta}, Elisabetta R. and {Misselt}, Karl and {Morse}, Jon A. and {Mulas}, Giacomo and {Neelamkodan}, Naslim and {Ohsawa}, Ryou and {Paladini}, Roberta and {Palumbo}, Maria Elisabetta and {Pathak}, Amit and {Pendleton}, Yvonne J. and {Petrignani}, Annemieke and {Pino}, Thomas and {Puga}, Elena and {Rangwala}, Naseem and {Rapacioli}, Mathias and {Ricca}, Alessandra and {Roman-Duval}, Julia and {Roser}, Joseph and {Roueff}, Evelyne and {Rouill{\'e}}, Ga{\"e}l and {Salama}, Farid and {Sales}, Dinalva A. and {Sandstrom}, Karin and {Sarre}, Peter and {Sciamma-O'Brien}, Ella and {Sellgren}, Kris and {Shenoy}, Sachindev S. and {Teyssier}, David and {Thomas}, Richard D. and {Togi}, Aditya and {Verstraete}, Laurent and {Witt}, Adolf N. and {Wootten}, Alwyn and {Ysard}, Nathalie and {Zettergren}, Henning and {Zhang}, Yong and {Zhang}, Ziwei E. and {Zhen}, Junfeng},
        title = "{PDRs4All: III. JWST's NIR spectroscopic view of the Orion Bar}",
      journal = {\aap},
         year = 2024,
        month = may,
       volume = {685},
          eid = {A74},
        pages = {A74},
          doi = {10.1051/0004-6361/202348244},
archivePrefix = {arXiv},
       eprint = {2310.08720},
 primaryClass = {astro-ph.GA},
       adsurl = {https://ui.adsabs.harvard.edu/abs/2024A&A...685A..74P}
}

@ARTICLE{Ricca2012,
       author = {{Ricca}, Alessandra and {Bauschlicher}, Charles W., Jr. and {Boersma},
        Christiaan and {Tielens}, Alexander G.~G.~M. and {Allamandola},
        Louis J.},
        title = "{The Infrared Spectroscopy of Compact Polycyclic Aromatic Hydrocarbons
        Containing up to 384 Carbons}",
      journal = {\apj},
         year = 2012,
        month = Jul,
       volume = {754},
          eid = {75},
        pages = {75},
          doi = {10.1088/0004-637X/754/1/75},
       adsurl = {https://ui.adsabs.harvard.edu/#abs/2012ApJ...754...75R}
}

@ARTICLE{Sandstrom2012,
   author = {{Sandstrom}, K.~M. and {Bolatto}, A.~D. and {Bot}, C. and {Draine}, B.~T. and 
	{Ingalls}, J.~G. and {Israel}, F.~P. and {Jackson}, J.~M. and 
	{Leroy}, A.~K. and {Li}, A. and {Rubio}, M. and {Simon}, J.~D. and 
	{Smith}, J.~D.~T. and {Stanimirovi{\'c}}, S. and {Tielens}, A.~G.~G.~M. and 
	{van Loon}, J.~T.},
    title = "{The Spitzer Spectroscopic Survey of the Small Magellanic Cloud (S$^{4}$MC): Probing the Physical State of Polycyclic Aromatic Hydrocarbons in a Low-metallicity Environment}",
  journal = {\apj},
archivePrefix = "arXiv",
   eprint = {1109.0999},
 primaryClass = "astro-ph.CO",
     year = 2012,
    month = jan,
   volume = 744,
      eid = {20},
    pages = {20},
      doi = {10.1088/0004-637X/744/1/20},
   adsurl = {http://adsabs.harvard.edu/abs/2012ApJ...744...20S}
}

@ARTICLE{Schutte1993,
   author = {{Schutte}, W.~A. and {Tielens}, A.~G.~G.~M. and {Allamandola}, L.~J.
	},
    title = "{Theoretical modeling of the infrared fluorescence from interstellar polycyclic aromatic hydrocarbons}",
  journal = {\apj},
     year = 1993,
    month = sep,
   volume = 415,
    pages = {397-414},
      doi = {10.1086/173173},
   adsurl = {http://adsabs.harvard.edu/abs/1993ApJ...415..397S}
}

@INPROCEEDINGS{Shannon2018,
 publisher = {SciPy},
 booktitle = "Proceedings of the 17th Python in Science Conference",
     title = "{Organic Molecules in Space: Insights from the {NASA} Ames Molecular Database in the era of the James Webb Space Telescope}",
     pages = {99},
       doi = {10.25080/majora-4af1f417-00f},
      year = 2018,
    author = {Shannon, M.~J. and Boersma, C.}
}

@ARTICLE{Smith07b,
   author = {{Smith}, J.~D.~T. and {Draine}, B.~T. and {Dale}, D.~A. and 
	{Moustakas}, J. and {Kennicutt}, Jr., R.~C. and {Helou}, G. and 
	{Armus}, L. and {Roussel}, H. and {Sheth}, K. and {Bendo}, G.~J. and 
	{Buckalew}, B.~A. and {Calzetti}, D. and {Engelbracht}, C.~W. and 
	{Gordon}, K.~D. and {Hollenbach}, D.~J. and {Li}, A. and {Malhotra}, S. and 
	{Murphy}, E.~J. and {Walter}, F.},
    title = "{The Mid-Infrared Spectrum of Star-forming Galaxies: Global Properties of Polycyclic Aromatic Hydrocarbon Emission}",
  journal = {\apj},
   eprint = {astro-ph/0610913},
     year = 2007,
    month = feb,
   volume = 656,
    pages = {770-791},
      doi = {10.1086/510549},
   adsurl = {http://adsabs.harvard.edu/abs/2007ApJ...656..770S}
}

@ARTICLE{Stock2016,
	author = {{Stock}, D.~J. and {Choi}, W.~D.-Y. and {Moya}, L.~G.~V. and 
	{Otaguro}, J.~N. and {Sorkhou}, S. and {Allamandola}, L.~J. and 
	{Tielens}, A.~G.~G.~M. and {Peeters}, E.},
	title = "{Polycyclic Aromatic Hydrocarbon Emission in Spitzer/IRS Maps. I. Catalog and Simple Diagnostics}",
	journal = {\apj},
	archivePrefix = "arXiv",
	eprint = {1601.07906},
	year = 2016,
	month = mar,
	volume = 819,
	eid = {65},
	pages = {65},
	doi = {10.3847/0004-637X/819/1/65},
	adsurl = {http://adsabs.harvard.edu/abs/2016ApJ...819...65S}
}

@ARTICLE{Stock2017,
	author = {{Stock}, D.~J. and {Peeters}, E.},
	title = "{Polycyclic Aromatic Hydrocarbon Emission in Spitzer/IRS Maps. II. A Direct Link between Band Profiles and the Radiation Field Strength}",
	journal = {\apj},
	archivePrefix = "arXiv",
	eprint = {1702.02691},
	year = 2017,
	month = mar,
	volume = 837,
	eid = {129},
	pages = {129},
	doi = {10.3847/1538-4357/aa5f54},
	adsurl = {http://adsabs.harvard.edu/abs/2017ApJ...837..129S}
}

@ARTICLE{Maltseva2018,
       author = {{Maltseva}, Elena and {Mackie}, Cameron J. and {Candian}, Alessandra and {Petrignani}, Annemieke and {Huang}, Xinchuan and {Lee}, Timothy J. and {Tielens}, Alexander G.~G.~M. and {Oomens}, Jos and {Buma}, Wybren Jan},
        title = "{High-resolution IR absorption spectroscopy of polycyclic aromatic hydrocarbons in the 3 {\ensuremath{\mu}}m region: role of hydrogenation and alkylation}",
      journal = {\aap},
         year = 2018,
        month = mar,
       volume = {610},
          eid = {A65},
        pages = {A65},
          doi = {10.1051/0004-6361/201732102},
archivePrefix = {arXiv},
       eprint = {1802.01497},
 primaryClass = {astro-ph.GA},
       adsurl = {https://ui.adsabs.harvard.edu/abs/2018A&A...610A..65M}
}

@ARTICLE{Mackie2016,
       author = {{Mackie}, Cameron J. and {Candian}, Alessandra and {Huang}, Xinchuan and
         {Maltseva}, Elena and {Petrignani}, Annemieke and {Oomens}, Jos and
         {Mattioda}, Andrew L. and {Buma}, Wybren Jan and {Lee}, Timothy J. and
         {Tielens}, Alexander G.~G.~M.},
        title = "{The anharmonic quartic force field infrared spectra of five non-linear polycyclic aromatic hydrocarbons: Benz[a]anthracene, chrysene, phenanthrene, pyrene, and triphenylene}",
      journal = {Journal of Chemical Physics},
         year = "2016",
        month = "Aug",
       volume = {145},
       number = {8},
          eid = {084313},
        pages = {084313},
          doi = {10.1063/1.4961438},
       adsurl = {https://ui.adsabs.harvard.edu/abs/2016JChPh.145h4313M}
}

@BOOK{Tielens2005,
       author = {{Tielens}, A.~G.~G.~M.},
       title = "{The Physics and Chemistry of the Interstellar Medium}",
         year = "2005",
         publisher = " Cambridge Univ. Press, Cambridge",
       adsurl = {https://ui.adsabs.harvard.edu/abs/2005pcim.book.....T}
}

@ARTICLE{Vicente2013,
       author = {{Vicente}, S. and {Bern{\'e}}, O. and {Tielens}, A.~G.~G.~M. and {Hu{\'e}lamo}, N. and {Pantin}, E. and {Kamp}, I. and {Carmona}, A.},
        title = "{Polycyclic Aromatic Hydrocarbon Emission in the Proplyd HST10: What is the Mechanism behind Photoevaporation?}",
      journal = {\apjl},
         year = 2013,
        month = mar,
       volume = {765},
       number = {2},
          eid = {L38},
        pages = {L38},
          doi = {10.1088/2041-8205/765/2/L38},
archivePrefix = {arXiv},
       eprint = {1302.0706},
 primaryClass = {astro-ph.SR},
       adsurl = {https://ui.adsabs.harvard.edu/abs/2013ApJ...765L..38V}
}

@ARTICLE{Werner2004,
       author = {{Werner}, M.~W. and {Uchida}, K.~I. and {Sellgren}, K. and {Marengo}, M. and {Gordon}, K.~D. and {Morris}, P.~W. and {Houck}, J.~R. and {Stansberry}, J.~A.},
        title = "{New Infrared Emission Features and Spectral Variations in NGC 7023}",
      journal = {\apjs},
         year = 2004,
        month = sep,
       volume = {154},
       number = {1},
        pages = {309-314},
          doi = {10.1086/422413},
archivePrefix = {arXiv},
       eprint = {astro-ph/0407213},
 primaryClass = {astro-ph},
       adsurl = {https://ui.adsabs.harvard.edu/abs/2004ApJS..154..309W}
}

@ARTICLE{Zang2022,
       author = {{Zang}, Rong Xuan and {Maragkoudakis}, Alexandros and {Peeters}, Els},
        title = "{The spatially resolved PAH characteristics in the Whirlpool Galaxy (M51a)}",
      journal = {\mnras},
         year = 2022,
        month = apr,
       volume = {511},
       number = {4},
        pages = {5142-5157},
          doi = {10.1093/mnras/stac214},
archivePrefix = {arXiv},
       eprint = {2201.11713},
 primaryClass = {astro-ph.GA},
       adsurl = {https://ui.adsabs.harvard.edu/abs/2022MNRAS.511.5142Z}
}

@ARTICLE{Lai2022,
       author = {{Lai}, Thomas S. -Y. and {Armus}, Lee and {U}, Vivian and {D{\'\i}az-Santos}, Tanio and {Larson}, Kirsten L. and {Evans}, Aaron and {Malkan}, Matthew A. and {Appleton}, Philip and {Rich}, Jeff and {M{\"u}ller-S{\'a}nchez}, Francisco and {Inami}, Hanae and {Bohn}, Thomas and {McKinney}, Jed and {Finnerty}, Luke and {Law}, David R. and {Linden}, Sean T. and {Medling}, Anne M. and {Privon}, George C. and {Song}, Yiqing and {Stierwalt}, Sabrina and {van der Werf}, Paul P. and {Barcos-Mu{\~n}oz}, Loreto and {Smith}, J.~D.~T. and {Togi}, Aditya and {Aalto}, Susanne and {B{\"o}ker}, Torsten and {Charmandaris}, Vassilis and {Howell}, Justin and {Iwasawa}, Kazushi and {Kemper}, Francisca and {Mazzarella}, Joseph M. and {Murphy}, Eric J. and {Brown}, Michael J.~I. and {Hayward}, Christopher C. and {Marshall}, Jason and {Sanders}, David and {Surace}, Jason},
        title = "{GOALS-JWST: Tracing AGN Feedback on the Star-forming Interstellar Medium in NGC 7469}",
      journal = {\apjl},
         year = 2022,
        month = dec,
       volume = {941},
       number = {2},
          eid = {L36},
        pages = {L36},
          doi = {10.3847/2041-8213/ac9ebf},
archivePrefix = {arXiv},
       eprint = {2209.06741},
 primaryClass = {astro-ph.GA},
       adsurl = {https://ui.adsabs.harvard.edu/abs/2022ApJ...941L..36L}
}

@ARTICLE{Lai2023,
       author = {{Lai}, Thomas S. -Y. and {Armus}, Lee and {Bianchin}, Marina and {D{\'\i}az-Santos}, Tanio and {Linden}, Sean T. and {Privon}, George C. and {Inami}, Hanae and {U}, Vivian and {Bohn}, Thomas and {Evans}, Aaron S. and {Larson}, Kirsten L. and {Hensley}, Brandon S. and {Smith}, J. -D.~T. and {Malkan}, Matthew A. and {Song}, Yiqing and {Stierwalt}, Sabrina and {van der Werf}, Paul P. and {McKinney}, Jed and {Aalto}, Susanne and {Buiten}, Victorine A. and {Rich}, Jeff and {Charmandaris}, Vassilis and {Appleton}, Philip and {Barcos-Mu{\~n}oz}, Loreto and {B{\"o}ker}, Torsten and {Finnerty}, Luke and {Kader}, Justin A. and {Law}, David R. and {Medling}, Anne M. and {Brown}, Michael J.~I. and {Hayward}, Christopher C. and {Howell}, Justin and {Iwasawa}, Kazushi and {Kemper}, Francisca and {Marshall}, Jason and {Mazzarella}, Joseph M. and {M{\"u}ller-S{\'a}nchez}, Francisco and {Murphy}, Eric J. and {Sanders}, David and {Surace}, Jason},
        title = "{GOALS-JWST: Small Neutral Grains and Enhanced 3.3 {\ensuremath{\mu}}m PAH Emission in the Seyfert Galaxy NGC 7469}",
      journal = {\apjl},
         year = 2023,
        month = nov,
       volume = {957},
       number = {2},
          eid = {L26},
        pages = {L26},
          doi = {10.3847/2041-8213/ad0387},
archivePrefix = {arXiv},
       eprint = {2307.15169},
 primaryClass = {astro-ph.GA},
       adsurl = {https://ui.adsabs.harvard.edu/abs/2023ApJ...957L..26L}
}

@ARTICLE{Garcia-Bernete2022,
       author = {{Garc{\'\i}a-Bernete}, I. and {Rigopoulou}, D. and {Alonso-Herrero}, A. and {Donnan}, F.~R. and {Roche}, P.~F. and {Pereira-Santaella}, M. and {Labiano}, A. and {Peralta de Arriba}, L. and {Izumi}, T. and {Ramos Almeida}, C. and {Shimizu}, T. and {H{\"o}nig}, S. and {Garc{\'\i}a-Burillo}, S. and {Rosario}, D.~J. and {Ward}, M.~J. and {Bellocchi}, E. and {Hicks}, E.~K.~S. and {Fuller}, L. and {Packham}, C.},
        title = "{A high angular resolution view of the PAH emission in Seyfert galaxies using JWST/MRS data}",
      journal = {\aap},
         year = 2022,
        month = oct,
       volume = {666},
          eid = {L5},
        pages = {L5},
          doi = {10.1051/0004-6361/202244806},
archivePrefix = {arXiv},
       eprint = {2208.11620},
 primaryClass = {astro-ph.GA},
       adsurl = {https://ui.adsabs.harvard.edu/abs/2022A&A...666L...5G}
}

@ARTICLE{Chastenet2023,
       author = {{Chastenet}, J{\'e}r{\'e}my and {Sutter}, Jessica and {Sandstrom}, Karin and {Belfiore}, Francesco and {Egorov}, Oleg V. and {Larson}, Kirsten L. and {Leroy}, Adam K. and {Liu}, Daizhong and {Rosolowsky}, Erik and {Thilker}, David A. and {Watkins}, Elizabeth J. and {Williams}, Thomas G. and {Barnes}, Ashley. T. and {Bigiel}, F. and {Boquien}, M{\'e}d{\'e}ric and {Chevance}, M{\'e}lanie and {Dale}, Daniel A. and {Kruijssen}, J.~M. Diederik and {Emsellem}, Eric and {Grasha}, Kathryn and {Groves}, Brent and {Hassani}, Hamid and {Hughes}, Annie and {Kreckel}, Kathryn and {Meidt}, Sharon E. and {Pan}, Hsi-An and {Querejeta}, Miguel and {Schinnerer}, Eva and {Whitcomb}, Cory M.},
        title = "{PHANGS-JWST First Results: Measuring Polycyclic Aromatic Hydrocarbon Properties across the Multiphase Interstellar Medium}",
      journal = {\apjl},
         year = 2023,
        month = feb,
       volume = {944},
       number = {2},
          eid = {L12},
        pages = {L12},
          doi = {10.3847/2041-8213/acac94},
       adsurl = {https://ui.adsabs.harvard.edu/abs/2023ApJ...944L..12C}
}

@ARTICLE{Egorov2023,
       author = {{Egorov}, Oleg V. and {Kreckel}, Kathryn and {Sandstrom}, Karin M. and {Leroy}, Adam K. and {Glover}, Simon C.~O. and {Groves}, Brent and {Kruijssen}, J.~M. Diederik and {Barnes}, Ashley. T. and {Belfiore}, Francesco and {Bigiel}, F. and {Blanc}, Guillermo A. and {Boquien}, M{\'e}d{\'e}ric and {Cao}, Yixian and {Chastenet}, J{\'e}r{\'e}my and {Chevance}, M{\'e}lanie and {Congiu}, Enrico and {Dale}, Daniel A. and {Emsellem}, Eric and {Grasha}, Kathryn and {Klessen}, Ralf S. and {Larson}, Kirsten L. and {Liu}, Daizhong and {Murphy}, Eric J. and {Pan}, Hsi-An and {Pessa}, Ismael and {Pety}, J{\'e}r{\^o}me and {Rosolowsky}, Erik and {Scheuermann}, Fabian and {Schinnerer}, Eva and {Sutter}, Jessica and {Thilker}, David A. and {Watkins}, Elizabeth J. and {Williams}, Thomas G.},
        title = "{PHANGS-JWST First Results: Destruction of the PAH Molecules in H II Regions Probed by JWST and MUSE}",
      journal = {\apjl},
         year = 2023,
        month = feb,
       volume = {944},
       number = {2},
          eid = {L16},
        pages = {L16},
          doi = {10.3847/2041-8213/acac92},
archivePrefix = {arXiv},
       eprint = {2212.09159},
 primaryClass = {astro-ph.GA},
       adsurl = {https://ui.adsabs.harvard.edu/abs/2023ApJ...944L..16E}
}

@ARTICLE{Rigopoulou2024,
       author = {{Rigopoulou}, Dimitra and {Donnan}, Fergus R. and {Garc{\'\i}a-Bernete}, Ismael and {Pereira-Santaella}, Miguel and {Alonso-Herrero}, Almudena and {Davies}, Ric and {Hunt}, Leslie K. and {Roche}, Patrick F. and {Shimizu}, Taro},
        title = "{Polycyclic Aromatic Hydrocarbon Emission in Galaxies as seen with JWST}",
      journal = {arXiv e-prints},
         year = 2024,
        month = jun,
          eid = {arXiv:2406.11415},
        pages = {arXiv:2406.11415},
          doi = {10.48550/arXiv.2406.11415},
archivePrefix = {arXiv},
       eprint = {2406.11415},
 primaryClass = {astro-ph.GA},
       adsurl = {https://ui.adsabs.harvard.edu/abs/2024arXiv240611415R}
}

@ARTICLE{Habing1968,
       author = {{Habing}, H.~J.},
        title = "{The interstellar radiation density between 912 A and 2400 A}",
      journal = {\bain},
         year = 1968,
        month = jan,
       volume = {19},
        pages = {421},
       adsurl = {https://ui.adsabs.harvard.edu/abs/1968BAN....19..421H}
}

@ARTICLE{Pilleri2012,
       author = {{Pilleri}, P. and {Montillaud}, J. and {Bern{\'e}}, O. and {Joblin}, C.},
        title = "{Evaporating very small grains as tracers of the UV radiation field in photo-dissociation regions}",
      journal = {\aap},
         year = 2012,
        month = jun,
       volume = {542},
          eid = {A69},
        pages = {A69},
          doi = {10.1051/0004-6361/201015915},
archivePrefix = {arXiv},
       eprint = {1204.4669},
 primaryClass = {astro-ph.GA},
       adsurl = {https://ui.adsabs.harvard.edu/abs/2012A&A...542A..69P}
}

@ARTICLE{Pilleri2015,
       author = {{Pilleri}, P. and {Joblin}, C. and {Boulanger}, F. and {Onaka}, T.},
        title = "{Mixed aliphatic and aromatic composition of evaporating very small grains in NGC 7023 revealed by the 3.4/3.3 {\ensuremath{\mu}}m ratio}",
      journal = {\aap},
         year = 2015,
        month = may,
       volume = {577},
          eid = {A16},
        pages = {A16},
          doi = {10.1051/0004-6361/201425590},
archivePrefix = {arXiv},
       eprint = {1502.04941},
 primaryClass = {astro-ph.GA},
       adsurl = {https://ui.adsabs.harvard.edu/abs/2015A&A...577A..16P}
}

@ARTICLE{Joblin1996,
       author = {{Joblin}, C. and {Tielens}, A.~G.~G.~M. and {Allamandola}, L.~J. and {Geballe}, T.~R.},
        title = "{Spatial Variation of the 3.29 and 3.40 Micron Emission Bands within Reflection Nebulae and the Photochemical Evolution of Methylated Polycyclic Aromatic Hydrocarbons}",
      journal = {\apj},
         year = 1996,
        month = feb,
       volume = {458},
        pages = {610},
          doi = {10.1086/176843},
       adsurl = {https://ui.adsabs.harvard.edu/abs/1996ApJ...458..610J}
}

@ARTICLE{Joblin2018,
       author = {{Joblin}, C. and {Bron}, E. and {Pinto}, C. and {Pilleri}, P. and {Le Petit}, F. and {Gerin}, M. and {Le Bourlot}, J. and {Fuente}, A. and {Berne}, O. and {Goicoechea}, J.~R. and {Habart}, E. and {K{\"o}hler}, M. and {Teyssier}, D. and {Nagy}, Z. and {Montillaud}, J. and {Vastel}, C. and {Cernicharo}, J. and {R{\"o}llig}, M. and {Ossenkopf-Okada}, V. and {Bergin}, E.~A.},
        title = "{Structure of photodissociation fronts in star-forming regions revealed by Herschel observations of high-J CO emission lines}",
      journal = {\aap},
         year = 2018,
        month = jul,
       volume = {615},
          eid = {A129},
        pages = {A129},
          doi = {10.1051/0004-6361/201832611},
archivePrefix = {arXiv},
       eprint = {1801.03893},
 primaryClass = {astro-ph.GA},
       adsurl = {https://ui.adsabs.harvard.edu/abs/2018A&A...615A.129J}
}

@ARTICLE{Hogerheijde1995,
       author = {{Hogerheijde}, Michiel R. and {Jansen}, David J. and {van Dishoeck}, Ewine F.},
        title = "{Millimeter and submillimeter observations of the Orion Bar. I. Physical structure.}",
      journal = {\aap},
         year = 1995,
        month = feb,
       volume = {294},
        pages = {792-810},
       adsurl = {https://ui.adsabs.harvard.edu/abs/1995A&A...294..792H}
}

@ARTICLE{Tielens1993,
       author = {{Tielens}, A.~G.~G.~M. and {Meixner}, M.~M. and {van der Werf}, P.~P. and {Bregman}, J. and {Tauber}, J.~A. and {Stutzki}, J. and {Rank}, D.},
        title = "{Anatomy of the Photodissociation Region in the Orion Bar}",
      journal = {Science},
         year = 1993,
        month = oct,
       volume = {262},
       number = {5130},
        pages = {86-89},
          doi = {10.1126/science.262.5130.86},
       adsurl = {https://ui.adsabs.harvard.edu/abs/1993Sci...262...86T}
}

@ARTICLE{Kassis2006,
       author = {{Kassis}, Marc and {Adams}, Joseph D. and {Campbell}, Murray F. and {Deutsch}, Lynne K. and {Hora}, Joseph L. and {Jackson}, James M. and {Tollestrup}, Eric V.},
        title = "{Mid-Infrared Emission at Photodissociation Regions in the Orion Nebula}",
      journal = {\apj},
         year = 2006,
        month = feb,
       volume = {637},
       number = {2},
        pages = {823-837},
          doi = {10.1086/498404},
       adsurl = {https://ui.adsabs.harvard.edu/abs/2006ApJ...637..823K}
}

@ARTICLE{Goicoechea2016,
       author = {{Goicoechea}, Javier R. and {Pety}, J{\'e}r{\^o}me and {Cuadrado}, Sara and {Cernicharo}, Jos{\'e} and {Chapillon}, Edwige and {Fuente}, Asunci{\'o}n and {Gerin}, Maryvonne and {Joblin}, Christine and {Marcelino}, Nuria and {Pilleri}, Paolo},
        title = "{Compression and ablation of the photo-irradiated molecular cloud the Orion Bar}",
      journal = {\nat},
         year = 2016,
        month = sep,
       volume = {537},
       number = {7619},
        pages = {207-209},
          doi = {10.1038/nature18957},
archivePrefix = {arXiv},
       eprint = {1608.06173},
 primaryClass = {astro-ph.GA},
       adsurl = {https://ui.adsabs.harvard.edu/abs/2016Natur.537..207G}
}

@ARTICLE{Esposito2024,
       author = {{Esposito}, Vincent J. and {Allamandola}, Louis J. and {Boersma}, Christiaan and {Bregman}, Jesse D. and {Fortenberry}, Ryan C. and {Maragkoudakis}, Alexandros and {Temi}, Pasquale},
        title = "{Anharmonic IR absorption spectra of the prototypical interstellar PAHs phenanthrene, pyrene, and pentacene in their neutral and cation states}",
      journal = {Molecular Physics},
         year = 2024,
        month = apr,
       volume = {122},
       number = {7-8},
          eid = {e2252936},
        pages = {e2252936},
          doi = {10.1080/00268976.2023.2252936},
       adsurl = {https://ui.adsabs.harvard.edu/abs/2024MolPh.12252936E}
}

@ARTICLE{Goicoechea2025,
       author = {{Goicoechea}, J.~R. and {Pety}, J. and {Cuadrado}, S. and {Bern{\'e}}, O. and {Dartois}, E. and {Gerin}, M. and {Joblin}, C. and {K{\l}os}, J. and {Lique}, F. and {Onaka}, T. and {Peeters}, E. and {Tielens}, A.~G.~G.~M. and {Alarc{\'o}n}, F. and {Bron}, E. and {Cami}, J. and {Canin}, A. and {Chapillon}, E. and {Chown}, R. and {Fuente}, A. and {Habart}, E. and {Kannavou}, O. and {Le Petit}, F. and {Santa-Maria}, M.~G. and {Schroetter}, I. and {Sidhu}, A. and {Trahin}, B. and {Van De Putte}, D. and {Zannese}, M.},
        title = "{PDRs4All: XII. Far-ultraviolet-driven formation of simple hydrocarbon radicals and their relation with polycyclic aromatic hydrocarbons}",
      journal = {\aap},
         year = 2025,
        month = apr,
       volume = {696},
          eid = {A100},
        pages = {A100},
          doi = {10.1051/0004-6361/202453350},
archivePrefix = {arXiv},
       eprint = {2503.03353},
 primaryClass = {astro-ph.GA},
       adsurl = {https://ui.adsabs.harvard.edu/abs/2025A&A...696A.100G}
}

@ARTICLE{Jochims1994,
       author = {{Jochims}, H.~W. and {Ruhl}, E. and {Baumgartel}, H. and {Tobita}, S. and {Leach}, S.},
        title = "{Size Effects on Dissociation Rates of Polycyclic Aromatic Hydrocarbon Cations: Laboratory Studies and Astrophysical Implications}",
      journal = {\apj},
         year = 1994,
        month = jan,
       volume = {420},
        pages = {307},
          doi = {10.1086/173560},
       adsurl = {https://ui.adsabs.harvard.edu/abs/1994ApJ...420..307J}
}

@ARTICLE{Ekern1998,
       author = {{Ekern}, Scott P. and {Marshall}, Alan G. and {Szczepanski}, Jan and {Vala}, Martin},
        title = "{Photodissociation of Gas-Phase Polycylic Aromatic Hydrocarbon Cations}",
      journal = {Journal of Physical Chemistry A},
         year = 1998,
        month = may,
       volume = {102},
       number = {20},
        pages = {3498-3504},
          doi = {10.1021/jp980488e},
       adsurl = {https://ui.adsabs.harvard.edu/abs/1998JPCA..102.3498E}
}

@ARTICLE{Zhen2015,
       author = {{Zhen}, Junfeng and {Castellanos}, Pablo and {Paardekooper}, Daniel M. and {Ligterink}, Niels and {Linnartz}, Harold and {Nahon}, Laurent and {Joblin}, Christine and {Tielens}, Alexander G.~G.~M.},
        title = "{Laboratory Photo-chemistry of PAHs: Ionization versus Fragmentation}",
      journal = {\apjl},
         year = 2015,
        month = may,
       volume = {804},
       number = {1},
          eid = {L7},
        pages = {L7},
          doi = {10.1088/2041-8205/804/1/L7},
archivePrefix = {arXiv},
       eprint = {1505.00576},
 primaryClass = {astro-ph.IM},
       adsurl = {https://ui.adsabs.harvard.edu/abs/2015ApJ...804L...7Z}
}

@ARTICLE{Cheng2011,
       author = {{Cheng}, Bing-Ming and {Chen}, Hui-Fen and {Lu}, Hsiao-Chi and {Chen}, Hong-Kai and {Alam}, M.~S. and {Chou}, Sheng-Lung and {Lin}, Meng-Yeh},
        title = "{Absorption Cross Section of Gaseous Acetylene at 85 K in the Wavelength Range 110-155 nm}",
      journal = {\apjs},
         year = 2011,
        month = sep,
       volume = {196},
       number = {1},
          eid = {3},
        pages = {3},
          doi = {10.1088/0067-0049/196/1/3},
       adsurl = {https://ui.adsabs.harvard.edu/abs/2011ApJS..196....3C}
}

@ARTICLE{Heays2017,
       author = {{Heays}, A.~N. and {Bosman}, A.~D. and {van Dishoeck}, E.~F.},
        title = "{Photodissociation and photoionisation of atoms and molecules of astrophysical interest}",
      journal = {\aap},
         year = 2017,
        month = jun,
       volume = {602},
          eid = {A105},
        pages = {A105},
          doi = {10.1051/0004-6361/201628742},
archivePrefix = {arXiv},
       eprint = {1701.04459},
 primaryClass = {astro-ph.SR},
       adsurl = {https://ui.adsabs.harvard.edu/abs/2017A&A...602A.105H}
}

@ARTICLE{Parikka2018,
       author = {{Parikka}, A. and {Habart}, E. and {Bernard-Salas}, J. and {K{\"o}hler}, M. and {Abergel}, A.},
        title = "{High-J CO emission spatial distribution and excitation in the Orion Bar}",
      journal = {\aap},
         year = 2018,
        month = sep,
       volume = {617},
          eid = {A77},
        pages = {A77},
          doi = {10.1051/0004-6361/201731975},
       adsurl = {https://ui.adsabs.harvard.edu/abs/2018A&A...617A..77P}
}

@ARTICLE{Bernard-Salas2012,
       author = {{Bernard-Salas}, J. and {Habart}, E. and {Arab}, H. and {Abergel}, A. and {Dartois}, E. and {Martin}, P. and {Bontemps}, S. and {Joblin}, C. and {White}, G.~J. and {Bernard}, J. -P. and {Naylor}, D.},
        title = "{Spatial variation of the cooling lines in the Orion Bar from Herschel/PACS}",
      journal = {\aap},
         year = 2012,
        month = feb,
       volume = {538},
          eid = {A37},
        pages = {A37},
          doi = {10.1051/0004-6361/201118083},
archivePrefix = {arXiv},
       eprint = {1111.3653},
 primaryClass = {astro-ph.SR},
       adsurl = {https://ui.adsabs.harvard.edu/abs/2012A&A...538A..37B}
}
